\documentclass[11pt,prb,onecolumn,showpacs,amssymb,superscriptaddress]{revtex4-2}
\usepackage{bbm}
\usepackage{graphicx}
\usepackage{amsmath}
\usepackage{amsfonts}
\usepackage{amsthm}
\usepackage{amssymb}
\usepackage{amsbsy}
\usepackage{wasysym}
\usepackage{bm}
\usepackage{bbm}
\usepackage{mathrsfs}
\usepackage{color}
\usepackage{hyperref}
\usepackage{braket}
\usepackage{soul}

\def\Xint#1{\mathchoice
   {\XXint\displaystyle\textstyle{#1}}%
   {\XXint\textstyle\scriptstyle{#1}}%
   {\XXint\scriptstyle\scriptscriptstyle{#1}}%
   {\XXint\scriptscriptstyle\scriptscriptstyle{#1}}%
   \!\int}
\def\XXint#1#2#3{{\setbox0=\hbox{$#1{#2#3}{\int}$}
     \vcenter{\hbox{$#2#3$}}\kern-.5\wd0}}
\def\dashint{\Xint-}

\setcitestyle{square}

\date{\today}
\begin{document}

\title{Random-matrix models of monitored quantum circuits}

\author{Vir B. Bulchandani}

\affiliation{Department of Physics, Princeton University, Princeton, New Jersey 08544, USA}
\affiliation{Institut f{\"u}r Theoretische Physik, Leibniz Universit{\"a}t Hannover, Appelstra{\ss}e 2, 30167 Hannover, Germany}

\author{S. L. Sondhi}
\affiliation{Rudolf Peierls Centre for Theoretical Physics, University of Oxford, Oxford OX1 3PU, United Kingdom}

\author{J. T. Chalker}
\affiliation{Rudolf Peierls Centre for Theoretical Physics, University of Oxford, Oxford OX1 3PU, United Kingdom}

\begin{abstract}
We study the competition between Haar-random unitary dynamics and measurements for unstructured systems of qubits. For projective measurements, we derive various properties of the statistical ensemble of Kraus operators analytically, including the purification time and the distribution of Born probabilities. The latter generalizes the Porter-Thomas distribution for random unitary circuits to the monitored setting and is log-normal at long times. We also consider weak measurements that interpolate between identity quantum channels and projective measurements. In this setting, we derive an exactly solvable Fokker-Planck equation for the joint distribution of singular values of Kraus operators, analogous to the Dorokhov-Mello-Pereyra-Kumar (DMPK) equation modelling disordered quantum wires. We expect that the statistical properties of Kraus operators we have established for these simple systems will serve as a model for the entangling phase of monitored quantum systems more generally.
\end{abstract}

\maketitle
\tableofcontents

\section{Introduction}
If a quantum computer is subjected to excessive random noise, it will cease to provide an advantage over classical devices. The resulting phase transition between ``quantum'' and ``classical'' computational regimes was first clearly identified by Aharonov~\cite{aharonov2000quantum}. Recently, this idea has enjoyed a resurgence in the context of the ``measurement-induced phase transition'' (MIPT) arising in quantum circuits subjected to random measurements~\cite{LiChenFisher,Chanetal,SkinnerRuhmanNahum,aletSchomerus,Choi_2020}. A well-studied model that realizes this transition is a chain of $L$ qubits acted upon by alternating unitary and measurement layers, consisting of local unitary gates and a spatial density $p>0$ of single-qubit projective measurements respectively. Under such dynamics, mixed initial states will undergo a process of ``dynamical purification'' whereby they tend to pure states at long times, with the R{\'e}nyi entropy of the system density matrix decaying to zero along generic quantum trajectories~\cite{Gullans_2020}.  The characteristic timescale for this process to occur is called the purification time, which we denote $\tau_{\mathrm{P}}$. For sufficiently small $p$, such systems realize an ``entangling'' phase, for which $\tau_{\mathrm{P}}$ is exponentially long in the system size, while for larger $p$, they realize a ``disentangling'' phase, for which $\tau_{\mathrm{P}}$ is independent of the system size. The two phases are separated by a critical point $p=p_c$ at which $\tau_{\mathrm{P}} \sim L$~\cite{Gullans_2020,Zabalo_2022}.

A particularly simple model for such dynamical purification is given by the ``monitored Haar-random quantum dot'', whose time evolution consists of Haar-random unitaries on the full $L$-qubit Hilbert space, i.e. unitary matrices drawn from the Haar measure on $U(N)$ where $N=2^L$, alternating with layers of independent single-qubit projective measurements on $pL$ of the qubits. Variants of this model were studied in Refs. \cite{fidkowski2021dynamical,nahum2021measurement,schomerus2022noisy}. These models realize only the entangling phase of the conventional MIPT, with a ``trivial'' critical point at $p_c = 1$. Nevertheless, we show below that these models yield a plethora of new and exact results, some of which should capture universal features of the entangling phase in spatially local systems, though we will not explicitly consider spatially local systems in this paper.

We achieve this by applying a combination of ideas from random-matrix theory and disorder physics to analyze this problem, which differ from the field-theoretic replica methods that have largely been the analytical tools of choice for analyzing monitored dynamical phases in previous work~\cite{MIPTReview}. This approach allows us to determine analytically the behaviour of various physical quantities that were previously mostly accessible by numerical simulations or by heuristic arguments, including the purification time, the dynamics of R{\'e}nyi entropies and the distribution of Born probabilities. We also introduce an unstructured model with near-identity weak measurements that can be seen as a monitored analogue of Dyson Brownian motion (see Refs. \cite{fidkowski2021dynamical} and \cite{schomerus2022noisy} for related constructions). The Fokker-Planck equation describing this model in the continuous time limit turns out to coincide with a known solvable model of Calogero-Sutherland type~\cite{Ipsen_2016}. This exact solution grants us a thorough understanding of the dynamics of the full set of singular values of Kraus operators along quantum trajectories, for all times.

Our perspective in this paper differs from most previous work on monitored quantum circuits in another important respect. Monitored dynamical phases and the transitions between them are usually diagnosed by averaging physical quantities over ensembles of quantum trajectories weighted by their Born probabilities, and the main objects of study are the numerical values of these averages~\cite{MIPTReview}. The quantities being averaged are furthermore usually nonlinear functions of the density matrix on each quantum trajectory, leading to a ``post-selection barrier'' that hinders direct comparison with experiments~\cite{Noel_2022,IBMExpt,GoogleExpt}. Post-selection and averaging tend to be advocated on the grounds that they are indispensable for probing ``typical'' quantum trajectories, which determine the monitored dynamical phase but are invisible at the level of the conventional density matrix. In this paper, we instead study time evolution along typical quantum trajectories directly (see also Ref. \cite{Zabalo_2022} and \cite{Piroli_2023}). Considering time evolution along single trajectories in lieu of Born-rule averaging can be motivated by analogy with the ergodic hypothesis in classical statistical physics, which similarly permits replacing ensemble averaging in chaotic systems with time-averaging along individual phase-space trajectories. However, in order to apply such reasoning to monitored quantum systems, we must first understand the extent to which distinct quantum trajectories are statistically alike. We must also understand the shape of the probability distributions that capture this statistical similarity. The latter will depend on both time and on the specific system under consideration, raising the question of how far such probability distributions reflect universal features of the underlying dynamical phase. To our knowledge, these probability distributions have not been studied in detail. Here, we derive them for monitored Haar-random quantum dots and return to questions of universality in the conclusion.

The paper is structured as follows. We first consider projective measurements and introduce the statistical ensembles of Kraus operators that will be the central objects of study. We present exact expressions for the Lyapunov spectrum of these Kraus operators, based on a mapping to so-called ``truncated unitary ensembles'' in random matrix theory~\cite{zyczkowski2000truncations,Forrester_2015}, from which we determine the purification time analytically. We then derive the exact distribution of Born probabilities, and explain how this generalizes the so-called Porter-Thomas distribution~\cite{porter1956fluctuations,Boixo_2018,Bouland_2018} for random unitary circuits to a two-parameter family of distributions determined by both the measurement density $p$ and the circuit depth $t$. Finally, we consider time evolution starting from maximally mixed initial states, which clarifies earlier results~\cite{Gullans_2020,nahum2021measurement,Li_2021} on the time evolution of R{\'e}nyi entropies as reflecting a crossover in time from a narrow Wigner-semicircle-like to a broad (and to a first approximation log-normal) distribution of singular values of Kraus operators.

We next turn to the weakly measured case, for which we derive a DMPK-like equation~\cite{dorokhov1982transmission,mello1988macroscopic} that captures the time evolution of the joint distribution function of singular values of Kraus operators. The resulting Fokker-Planck equation is exactly solvable~\cite{Ipsen_2016} in a manner analogous to the time-reversal-symmetry-breaking case of the DMPK equation~\cite{beenakker1993nonlogarithmic}. Using this exact solution, we demonstrate explicitly that the joint distribution function of singular values exhibits a ``semicircle-to-square'' crossover between log-GUE statistics and log-normal statistics as a function of time. This represents a remarkably complete understanding of this model's dynamics that corroborates our results on projective measurements. We conjecture that 
this behaviour is universal for entangling phases of monitored quantum systems, in the same manner that random matrix theory captures 
the spectral properties of generic closed quantum systems.

\section{Projective measurements}
\subsection{Kraus operator ensembles}
\begin{figure}
    \centering
    \includegraphics[width = \linewidth]{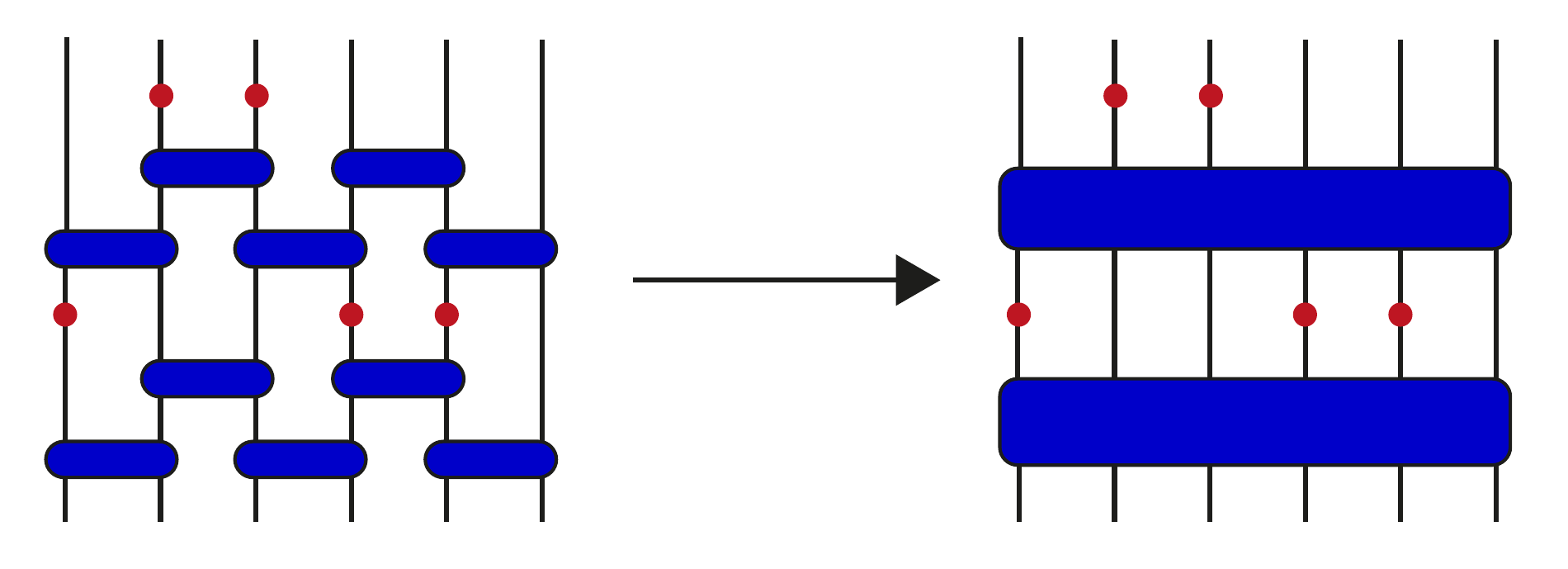}
    \caption{A schematic illustration of how the Haar-random monitored quantum circuits that we consider in this work (right) differ from more standard spatially local examples (left). The legs at the bottom of each picture correspond to individual qubits, blue rectangles depict Haar random unitary gates acting on the Hilbert space of the incoming legs, and red dots indicate single-qubit projective measurements.}
    \label{FigSchematic}
\end{figure}
Consider a quantum dot of $L$ qubits, acted upon by pairs of alternating unitary and measurement layers that together constitute individual time steps. Each unitary layer consists of a Haar-random unitary operator acting on all $L$ qubits and drawn from the Haar measure on $U(N)$, where $N=2^L$ throughout this paper. Each measurement layer comprises independent, projective, single-qubit $\hat{Z}$ measurements acting on a mean number of qubits $pL$ per layer, with $0 \leq p \leq 1$. The specific spatial probability distribution of these measurements will depend on the specific model of interest, to be fixed below. The difference between the models that we consider here and more standard~\cite{MIPTReview} spatially local Haar random quantum circuits is depicted schematically in Fig. \ref{FigSchematic}. We denote the string of measurement outcomes in the $j$th measurement layer by $\mathbf{m}_j$, and denote the measurement history of the whole circuit along a given quantum trajectory by the tuple $\mathbf{m}=(\mathbf{m}_1,\mathbf{m}_2,\ldots,\mathbf{m}_t)$. We denote the projection operator corresponding to projection onto the measurement outcome $\mathbf{m}_j$ by $\hat{P}_{\mathbf{m}_j}$. In general, the rank of $\hat{P}_{\mathbf{m}_j}$ equals $2^{L-|\mathbf{m}_{j}|}$, where $|\mathbf{m}_j|$ denotes the number of measurements made at time $j$.

For a pure initial state $|\psi(0)\rangle$, the time-evolved state along a given quantum trajectory $\mathbf{m}$ can be written as~\cite{MIPTReview}
\begin{equation}
|\psi(t)\rangle = \hat{K}_{\mathbf{m}}(t) |\psi(0)\rangle/\sqrt{p(\mathbf{m})},
\end{equation}
where the Kraus operator
\begin{equation}
\label{eq:Kraus}
\hat{K}_{\mathbf{m}}(t) = \hat{P}_{\mathbf{m}_t} \hat{U}_{t} \ldots \hat{P}_{\mathbf{m}_1} \hat{U}_1  
\end{equation}
and the Born probability of the string of measurement outcomes $\mathbf{m}$ is given by
\begin{equation}
p(\mathbf{m}) = \langle \psi(0) | \hat{K}_{\mathbf{m}}(t)^\dagger \hat{K}_{\mathbf{m}}(t) | \psi(0) \rangle.
\end{equation}

This can be generalized to arbitrary initial density matrices $\hat{\rho}(0)$, whose time evolution under this dynamics is given by
\begin{equation}
\label{eq:densitymatrix}
\hat{\rho}(t) = \sum_{\mathbf{m}} \hat{K}_{\mathbf{m}}(t) \hat{\rho}(0) \hat{K}_{\mathbf{m}}^\dagger(t).
\end{equation}
Where applicable, we make the choice~\cite{Gullans_2020} to unravel this evolution as an ensemble of single-trajectory density matrices
\begin{equation}
\hat{\rho}_{\mathbf{m}}(t) = \frac{\hat{K}_{\mathbf{m}}(t)\hat{\rho}(0)\hat{K}_{\mathbf{m}}^\dagger(t)}{p(\mathbf{m})}
\end{equation}
drawn with probabilities $p(\mathbf{m}) = \mathrm{Tr}[\hat{K}_{\mathbf{m}}(t)\hat{\rho}(0)\hat{K}_{\mathbf{m}}^\dagger(t)]$. We refer to the review article Ref. \cite{MIPTReview} for a more detailed discussion of Kraus operators for monitored quantum circuits. When considering mixed initial states, for simplicity we will always restrict our attention to the maximally mixed initial state
\begin{equation}
\label{eq:maxmix}
\hat{\rho}(0) = \frac{1}{2^L}\mathbbm{1},
\end{equation}
for which
\begin{equation}
\label{eq:maxmixevo}
\hat{\rho}_{\mathbf{m}}(t) = \frac{\hat{K}_{\mathbf{m}}(t)\hat{K}_{\mathbf{m}}^\dagger(t)}{\mathrm{Tr}[\hat{K}_{\mathbf{m}}(t)\hat{K}_{\mathbf{m}}^\dagger(t)]}
\end{equation}
along each quantum trajectory. 

The basic quantity of interest in this work is the statistical ensemble of Kraus operators $\{\hat{K}_{\mathbf{m}}\}$, defined in general by first sampling over the Haar measure, then sampling over measurement locations and finally sampling over measurement outcomes. Since we will be focusing on Haar random unstructured systems, the latter average will mostly be redundant, and expectation values $\mathbb{E}$ will always denote expectation values along a fixed quantum trajectory with respect to the Haar measure on each unitary layer unless specified otherwise. Thus we do not explicitly consider Born-rule averaged quantities in this work, although we address the distribution of Born probabilities with respect to the Haar measure in Section \ref{sec:born}.

The main tool at our disposal for studying the ensemble of Kraus operators $\hat{K}_{\mathbf{m}}$ will be their singular value decomposition, 
\begin{equation}
\label{eq:SVD}
\hat{K}_{\mathbf{m}}(t) = \hat{V}_t\hat{D}_t \hat{W}_t^\dagger
\end{equation}
where we write $\hat{D}_t = \mathrm{diag}(\sigma_1(t),\ldots,\sigma_N(t))$, with the convention that $\sigma_1(t) \geq \ldots \geq \sigma_N(t) \geq 0$. We emphasize that the matrices of singular vectors $\hat{V}_t,\, \hat{W}_t$ and the singular values $\sigma_n(t)$ are random variables that depend sensitively on the circuit realization, and suppress the measurement history in Eq. \eqref{eq:SVD} for notational convenience.

\subsection{Rank collapse and dynamical purification}
\label{sec:RankCollapseandModels}
Because the Kraus operators above generically include projective measurements, their rank $r(t) = \mathrm{rk}[\hat{K}_{\mathbf{m}}(t)]$ is a random variable that does not increase in time and $\sigma_n(t) = 0$ for $n > r(t)$. The decay of the rank defines a ``rank-collapse time'', given by the mean stopping time
\begin{equation}
\label{eq:stoppingtime}
\tau_{\mathrm{R.C.}} = \mathbb{E}[\min{\{t:r(t)=1\}}],
\end{equation} 
which is infinite if the rank does not decay.

In general, the singular values $\sigma_n(t)$ will decay in an average sense as $t \to \infty$; in particular, the Oseledets ergodic theorem~\cite{oseledets1968multiplicative} guarantees the existence of Lyapunov exponents
\begin{equation}
\lambda_n = \lim_{t \to \infty} \frac{\log{\sigma_n(t)}}{t}
\end{equation}
with probability one, provided that $n \leq r(t)$ with probability one for all time. To see the physical meaning of the first few singular values and their Lyapunov exponents, suppose that $\lambda_2$ is defined and consider time evolution from a maximally mixed state, as in Eq. \eqref{eq:maxmixevo}. Then
\begin{equation}
\hat{K}_{\mathbf{m}}(t) \hat{K}_{\mathbf{m}}(t)^\dagger \sim \sigma_1^2(t) \mathbf{v}_1 \mathbf{v}_1^\dagger + \sigma_2^2(t) \mathbf{v}_2 \mathbf{v}_2^\dagger, \quad t \to \infty,
\end{equation}
where $\mathbf{v}_n$ denotes the $n$th column of $\hat{V}_t$ in Eq. \eqref{eq:SVD}. Thus the Born probability
\begin{equation}
p(\mathbf{m}) \sim \frac{1}{N}\sigma_1^2(t), \quad t \to \infty,
\end{equation}
decays at a rate $\tau^{-1} = 2\lambda_1$~\cite{Zabalo_2022}, while the trajectory density matrix
\begin{equation}
\label{eq:purifiedstate}
\hat{\rho}_{\mathbf{m}}(t) \sim \mathbf{v}_1 \mathbf{v}_1^\dagger + \left(\frac{\sigma_2^2(t)}{\sigma_1^2(t)}\right) \mathbf{v}_2 \mathbf{v}_2^\dagger, \quad t \to \infty.
\end{equation}
It is clear that the decay of the ratio 
\begin{equation}
\label{eq:ratio}
\nu(t) = \frac{\sigma_2^2(t)}{\sigma_1^2(t)}
\end{equation} 
determines the rate of convergence of $\hat{\rho}_{\mathbf{m}}(t)$ to a pure state. We thus define the ``purification time'' $\tau_{\mathrm{P}}$ by
\begin{equation}
\label{eq:purificationtime}
\tau_{\mathrm{P}}^{-1} = \lim_{t\to\infty} \frac{-\mathbb{E}[\log{\nu(t)}]}{t} = 2(\lambda_1-\lambda_2).
\end{equation}
For more general initial states, such as pure states, $\tau_{\mathrm{P}}^{-1}$ is properly thought of as the expected rate of convergence of the trajectory density matrix to a rank-one projection along the random singular vector $\mathbf{v}_1$. The specific distribution of $\mathbf{v}_1$ will depend on the specific model and monitored dynamical phase under consideration. Thus a more general perspective on the purification time $\tau_{\mathrm{P}}$ is that it defines the timescale on which a monitored quantum system ``forgets'' its initial state~\cite{fidkowski2021dynamical} and begins to reveal universal properties of its monitored dynamical phase. We return to this point in the conclusion. 

The above discussion reveals that there are two important timescales that determine the fate of the monitored quantum dot (and indeed arbitrary projectively measured quantum circuits) at asymptotically long times, namely the rank-collapse time and the purification time. However, these two timescales are in tension, because standard results~\cite{oseledets1968multiplicative,benettin1980lyapunov} on the existence of Lyapunov exponents $\lambda_n$ for $n>1$ require that $\sigma_n(t)$ is generically non-zero for all time. Thus in order to define the purification time or higher Lyapunov exponents rigorously, we require that $\tau_{\mathrm{R.C.}}= \infty$. This is not the case for the standard formulation of monitored random circuits, according to which measurements are performed randomly and independently at every site~\cite{LiChenFisher} leading to rank collapse in finite time. Previous work on models with rank collapse implicitly assumes a parametric separation of scales
\begin{equation}
\label{eq:sepscales}
\tau_{\mathrm{P}} \ll \tau_{\mathrm{R.C.}}
\end{equation}
in $L$, and then estimates $\tau_{\mathrm{P}}$ numerically by simulating the system for times much shorter than $\tau_{\mathrm{R.C.}}$ (see~\cite{Zabalo_2022}). The downside of this approach is that there is an inherent ``fuzziness'' in the definition of $\tau_{\mathrm{P}}$ on the order of inevitable statistical fluctuations $1/\sqrt{\tau_{\mathrm{R.C.}}}$ induced by rank collapse. While this is mostly a mathematical subtlety when it comes to estimating the purification time numerically (since for the canonical spatially local model $\tau_{\mathrm{R.C.}} \sim 1/(2p)^L \gg 1$ by a mapping to bond percolation~\cite{SkinnerRuhmanNahum}), it becomes a serious obstacle even for numerical estimates of higher Lyapunov exponents $\lambda_n$ with $n > 2$, which succumb to rank reduction at much earlier times than $\tau_{\mathrm{R.C.}}$.

We resolve this subtlety for monitored quantum dots by noting that the timescales $\tau_{\mathrm{R.C.}}$ and $\tau_{\mathrm{P}}$ naturally pertain to two distinct microscopic models, with and without rank collapse respectively, for which the separation of scales Eq. \eqref{eq:sepscales} can be proved analytically as $L \to \infty$. This then justifies attempting to estimate $\tau_{\mathrm{P}}$ numerically in the model with rank collapse. These models, which we refer to as ``Model I'' and ``Model II'' respectively, differ only in the spatial distribution of measurements in each measurement layer and are defined as follows.

For Model I, the measurements in each measurement layer are performed randomly and independently at each qubit with probability $p$, as for the standard formulation of the MIPT~\cite{LiChenFisher}. Then the number of measurements in each layer fluctuates and rank collapse occurs when all the qubits in a given layer are measured, which occurs with probability $p^L$. For Model II, we perform a fixed number of measurements $pL \in \{0,1,\ldots,L\}$ per layer. Note that for the monitored quantum dots under consideration in this paper, the specific location of these measurements does not matter, by Haar randomness of the unitary layers.

For both models, we can determine the rank-collapse time in finite systems analytically, in contrast to the situation for spatially local models, for which closed forms are not easily obtained (even if the asymptotic behaviour is understood~\cite{SkinnerRuhmanNahum}). First consider Model I. From Eq. \eqref{eq:stoppingtime}, the rank-collapse time is the expected value of the stopping time $T$ such that $r(T)=1$ and $r(t) > 1$ for $t < T$. Note that $\mathbb{P}(T=t) = (1-p^L)^{t-1}p^L$. Thus
\begin{equation}
\tau_{\mathrm{R.C.,I}} = \mathbb{E}[T] = \sum_{t=1}^{\infty} t(1-p^L)^{t-1}p^L = \frac{1}{p^L}.
\end{equation}
Next consider Model II. In this case $r(t) = 2^{(1-p)L} > 1$ almost surely for $p <1$ and it follows that
\begin{equation}
\tau_{\mathrm{R.C.,II}} = \mathbb{E}[T] = \infty, \quad p <1.
\end{equation}
Below, we will restrict our attention to Model II with $0<p<1$ and write $M = 2^{(1-p)L}$ for the rank of its Kraus operators.

\subsection{Mapping to truncated unitary and Ginibre ensembles}
\label{sec:MaptoRMT}
The ``truncated unitary ensembles'' of random matrices~\cite{zyczkowski2000truncations} consist of square submatrices of Haar random matrices drawn from $U(N)$. We can relate the Kraus operators Eq. \eqref{eq:Kraus} for Model II to products of $M \times M$ truncated unitary matrices drawn from $U(N)$ as follows. We first note that by Haar randomness and transitivity of $U(N)$ on measurement outcomes, the statistics of singular values of $\{\hat{K}_{\mathbf{m}}(t)\}$ is unchanged if we fix the measurement outcomes $\hat{P}_{\mathbf{m}_j} = \hat{P}$ in each layer to be the same. Writing $\overset{\mathrm{s.v.}}{\sim}$ for equality of singular value statistics, we have
\begin{equation}
\hat{K}_{\mathbf{m}}(t) =  \hat{P}_{\mathbf{m}_t} \hat{U}_{t} \ldots \hat{P}_{\mathbf{m}_1} \hat{U}_1  \overset{\mathrm{s.v.}}{\sim} \hat{P} \hat{U}_t  \ldots \hat{P} \hat{U}_1.
\end{equation}
at each time step $t$. To proceed further, we use the fact that $\hat{P}^2=\hat{P}$ to write
\begin{equation}
\hat{P} \hat{U}_t  \ldots \hat{P} \hat{U}_1 = (\hat{P} \hat{U}_t \hat{P})(\hat{P}\hat{U}_{t-1}\hat{P}) \ldots (\hat{P} \hat{U}_2\hat{P}) \hat{P}\hat{U}_1,
\end{equation}
implying that
\begin{equation}
\label{eq:maptoTUE}
\hat{K}_{\mathbf{m}}(t) \overset{\mathrm{s.v.}}{\sim} \hat{R}_{t} \hat{R}_{t-1} \ldots \hat{R}_2 \hat{S}_1,
\end{equation}
where in the computational basis, $\hat{R}_j = \hat{P} \hat{U}_j \hat{P}$ is an $M$-by-$M$ truncated unitary matrix padded by $N-M$ rows and columns of zeros, with the same configuration of nonzero entries for each $j$, and $\hat{S}_1 = \hat{P}\hat{U_1}$ has $M$ unit singular values and $N-M$ zero singular values, since
\begin{equation}
\hat{S}_1\hat{S}_1^\dagger = \hat{P}.
\end{equation}
It follows from this analysis that the singular values of $\hat{K}_\mathbf{m}(t)$ are distributed as the singular values of products of $t-1$ truncated unitary matrices. While various properties of products of truncated unitary matrices have been derived analytically in the literature~\cite{Akemann_2014,Forrester_2015,Akemann_2015,ahn2022lyapunov}, including explicit expressions for their Lyapunov spectrum that we will discuss further below, the simplest results are obtained in the limit $M/N \to 0$, in which $\hat{K}$ is distributed as a product of Ginibre random matrices~\cite{Akemann_2014}. In the physical context of monitored random circuits introduced above, this regime is realized in the thermodynamic limit $L \to \infty$ in which the number of qubits tends to infinity at any fixed $p>0$.

Intuitively, the emergence of Ginibre ensembles corresponds to the fact that as $M/N \to 0$, the non-zero elements of $\hat{R}_j$ look ``random'', in the sense that they are asymptotically distributed as the elements of a Ginibre matrix, up to rescaling~\cite{petz2003asymptotics,Mastrodonato_2007}. To be precise, let $\hat{A}_j$ denote an $M$-by-$M$ complex Ginibre matrix whose elements are i.i.d. with probability distribution function $f(v) = \frac{1}{\pi} e^{-|v|^2}$ and define
\begin{equation}
\hat{B}_j = \frac{1}{\sqrt{N}} \hat{A}_j.
\end{equation}
This choice of normalization ensures that in the absence of truncation, $M=N$, the expected squared norm of each row and each column of $\hat{B}_j$ coincides with the expected squared norm of each row and each column of an $N$-by-$N$ unitary matrix. It can then be shown~\cite{Akemann_2014} that as $M/N \to 0$, the product of Ginibre random matrices
\begin{equation}
\label{eq:GinProd}
\hat{G}(t) = \hat{B}_{t-1} \hat{B}_{t-2} \ldots \hat{B}_1, \quad \hat{G}(1) = \mathbbm{1}_M,
\end{equation}
has the same probability distribution as the non-zero elements of $\hat{R}_t \hat{R}_{t-1} \ldots \hat{R}_2$, and therefore that the non-zero singular values of  $\hat{K}_{\mathbf{m}}(t)$ and $\hat{G}(t)$ are identically distributed in the thermodynamic limit. We note that the spectral density of the product $\hat{G}(t)$ has been studied in some detail~\cite{Akemann_2013,Akemann_2019}, and will make use of some of these results below.

\subsection{Computation of Lyapunov exponents and the purification time}
We now compute the entire Lyapunov spectrum for Model II. This follows immediately from the mapping Eq. \eqref{eq:maptoTUE} from Model II to the products of truncated unitary matrices, combined with an expression for the Lyapunov exponents of such products derived recently~\cite{Forrester_2015,Akemann_2015,ahn2022lyapunov}. Together, these results imply that the Lyapunov exponents for Model II are given by 
\begin{equation}
\label{eq:ExactLyapProj}
\lambda_n = -\frac{1}{2}(\psi(N-n+1) - \psi(M-n+1)), \quad n=1,2,\ldots,M,
\end{equation} 
where $\psi(x) = \Gamma'(x)/\Gamma(x)$ denotes the digamma function.

\begin{figure}[t]
    \centering
    \includegraphics[width= 0.75\linewidth]{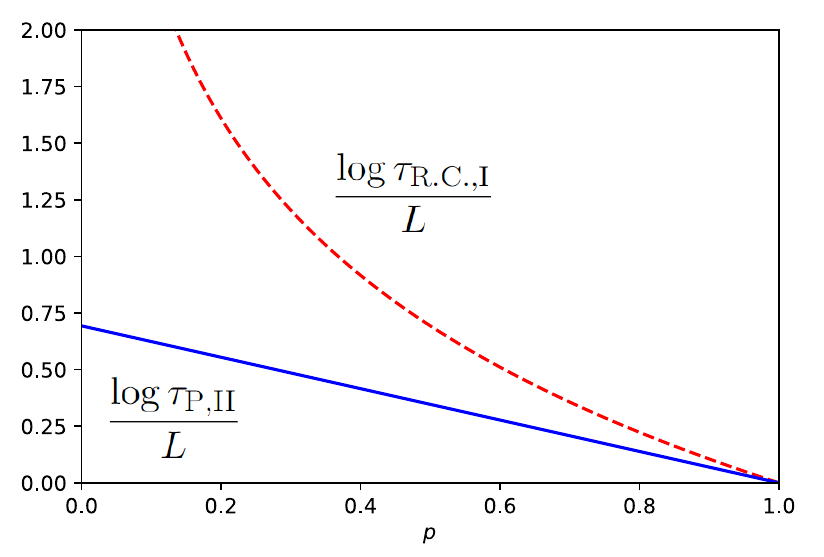}
    \caption{The rank-collapse time for Model I (dashed red line) versus the purification time for Model II (solid blue line). It is clear from this plot that these timescales are exponentially well separated as $L \to \infty$, except in the immediate vicinity of $p=1$.}
    \label{FigTimescales}
\end{figure}

It follows from Eq. \eqref{eq:ExactLyapProj} that the inverse purification time for Model II
\begin{equation}
\tau_{\mathrm{P,II}}^{-1} = 2(\lambda_1-\lambda_2) = \frac{1}{M-1}-\frac{1}{N-1}
\end{equation}
by the standard digamma function identity~\cite{abramowitz1968handbook}  $\psi(n+1) = \psi(n) + \frac{1}{n}$. We deduce that
\begin{equation}
\label{eq:inequality}
\lim_{L\to\infty} \frac{\log{\tau_{\mathrm{P,II}}}}{L} = (1-p)\log{2} < - \log{p} = \lim_{L\to\infty} \frac{\log{\tau_{\mathrm{R.C.,I}}}}{L}
\end{equation}
for $p < 1$, implying that
\begin{equation}
\label{eq:carefulsepscales}
\tau_{\mathrm{P,II}} \ll \tau_{\mathrm{R.C.,I}}
\end{equation}
for $L \gg 1$ and $p<1$, which is one way to make Eq. \eqref{eq:sepscales} precise. The functions of $p$ appearing on either side of Eq. \eqref{eq:inequality} are plotted in Fig. \ref{FigTimescales}; it is clear that they coincide only in the approach to the ``transition'' as $p \to 1^-$. In particular, we have proved that the purification time
\begin{equation}
\label{eq:taupM}
\tau_{\mathrm{P}} \sim M \sim 2^{(1-p)L}, \quad L \to \infty,
\end{equation}
grows exponentially in the system size, and having established Eq. \eqref{eq:sepscales} we henceforth suppress the subscript specifying Model II.

\subsection{The distribution of Born probabilities}
\label{sec:born}
Now consider applying Model II dynamics to an arbitrary initial pure state $|\psi\rangle$, with a view to determining the distribution of Born probabilities as a function of both the density of measurements $p$ and the circuit depth $t$. The Born probability of measuring a string of measurement outcomes $\mathbf{m}$ after time $t$ can be written explicitly as $
p(\mathbf{m}) = \| \hat{P}_{\mathbf{m}_t} \hat{U}_{t} \ldots \hat{P}_{\mathbf{m}_1} \hat{U}_1 |\psi \rangle \|^2$. Introducing rotated projection operators $\hat{P}'_{\mathbf{m}_j} = \hat{U}_1^\dagger \ldots \hat{U}_{j}^\dagger \hat{P}_{\mathbf{m}_j} \hat{U}_j \ldots \hat{U}_1$, we can write this as
\begin{equation}
p(\mathbf{m}) = \| \hat{P}'_{\mathbf{m}_t}\hat{P}'_{\mathbf{m}_{t-1}}\ldots \hat{P}'_{\mathbf{m}_1} |\psi \rangle \|^2.
\end{equation}
This implies that
\begin{equation}
\label{eq:Markov}
p(\mathbf{m}) = p(\mathbf{m}_t|\mathbf{m}_{t-1}\ldots \mathbf{m}_{1})p(\mathbf{m}_{t-1}|\mathbf{m}_{t-2}\ldots \mathbf{m}_{1})\ldots p(\mathbf{m}_{2}| \mathbf{m}_{1})p(\mathbf{m}_1),
\end{equation}
where
\begin{equation}
p(\mathbf{m}_j|\mathbf{m}_{j-1}\ldots \mathbf{m}_{1}) = \frac{\| \hat{P}'_{\mathbf{m}_j}\hat{P}'_{\mathbf{m}_{j-1}}\ldots \hat{P}'_{\mathbf{m}_1} |\psi \rangle \|^2}{\| \hat{P}'_{\mathbf{m}_{j-1}}\hat{P}'_{\mathbf{m}_{j-2}}\ldots \hat{P}'_{\mathbf{m}_1} |\psi \rangle \|^2}.
\end{equation}
Note that so far we have assumed nothing about the distribution of unitary layers $\hat{U}_j$. If we now assume that $\hat{U}_j$ are Haar-random matrices on the full Hilbert space, a drastic simplification occurs and the measurement outcomes at distinct time steps become pairwise uncorrelated. Explicitly, we have 
\begin{equation}
p(\mathbf{m}_j|\mathbf{m}_{j-1}\ldots \mathbf{m}_1) \sim \frac{\sum_{k=1}^{M} |v_{j,k}|^2}{\sum_{k=1}^{N} |v_{j,k}|^2}
\end{equation}
for all $j$, where the $v_{j,k}$ are i.i.d. complex Gaussian random variables with probability density function (p.d.f.) $f(v) = \frac{1}{\pi}e^{-|v|^2}$. Note that sums of $K$ such variables $\sum_{k=1}^K |v_{j,k}|^2$ are gamma distributed, with p.d.f. $f_K(x) = \frac{1}{\Gamma(K)}x^{K-1}e^{-x}$. Thus let $X_{K,j}$ denote a set of independently distributed gamma random variables, each with p.d.f. $f_K(x)$. Then 
\begin{equation}
\label{eq:marginaldist}
p(\mathbf{m}_j|\mathbf{m}_{j-1}\ldots \mathbf{m}_1) \sim \frac{X_{M,j}}{X_{M,j} + X_{N-M,j}} \sim \mathrm{Beta}(M,N-M)
\end{equation}
follows a beta distribution with parameters $M$ and $N-M$ and p.d.f 
\begin{equation}
\label{eq:betadist}
f_{M,N-M}(x) = \frac{\Gamma(N)}{\Gamma(M)\Gamma(N-M)}x^{M-1}(1-x)^{N-M-1}.
\end{equation}
The usual Porter-Thomas distribution for random unitary circuits~\cite{Boixo_2018} corresponds to measuring every qubit after each unitary layer, and is recovered from Eqs. \eqref{eq:marginaldist} and \eqref{eq:betadist} in the limit that $p=1$ and $M=1$, for which
\begin{equation}
\label{eq:PT}
p_{\mathrm{Porter-Thomas}}(\mathbf{m}_j|\mathbf{m}_{j-1}\ldots \mathbf{m}_1) \sim \mathrm{Beta}(1,N-1)
\end{equation}
is asymptotically exponentially distributed with mean $1/N$ for large $N$. 

Let us now return to the original problem of determining the probability distribution function of $p(\mathbf{m})$ for Model II. We showed above that
\begin{equation}
\label{eq:Borndist}
p(\mathbf{m}) \sim \prod_{j=1}^t Y_j, 
\end{equation}
for i.i.d. beta random variables $Y_j \sim \mathrm{Beta}(M,N-M)$. This implies in particular that 
\begin{equation}
\log p(\mathbf{m}) \sim \sum_{j=1}^t \log{Y_j},
\end{equation}
from which log-normality of $p(\mathbf{m})$ at long times is immediate. It follows by the central limit theorem that
\begin{equation}
\log p(\mathbf{m}) \to \mu t + \varsigma t^{1/2} Z_t,\quad t \to \infty,
\end{equation}
in distribution, where the mean
\begin{equation}
\label{eq:bornmean}
\mu = \psi(M)-\psi(N) < 0,
\end{equation}
the variance
\begin{equation}
\label{eq:bornvar}
\varsigma^2 = \psi'(M)-\psi'(N) > 0,
\end{equation}
and $Z_t \sim \mathcal{N}(0,1)$ is a unit normal random variable. Thus we have shown that the Born probabilities $p({\mathbf{m}})$ for Model II  are asymptotically log-normally distributed, with a shape that is determined by the parameters $N$, $M$ and $t$.

Exact expressions for the distributions of Born probabilities can be obtained as follows. Denoting the characteristic function of a given random variable $W$ by $\varphi_W(\theta) = \mathbb{E}[e^{i\theta W}]$, it follows by Eq. \eqref{eq:Borndist} that
\begin{equation}
\varphi_{\log{p(\mathbf{m})}}(\theta) = \mathbb{E}[e^{i\theta \sum_{j=1}^t \log{Y_j}}] = \prod _{j=1}^t \mathbb{E}[e^{i\theta \log{Y}_j}] = [\varphi_{\log{Y_1}}(\theta)]^t.
\end{equation}
The characteristic function of the beta distribution Eq. \eqref{eq:betadist} is given by
\begin{equation}
\varphi_{\log{Y_1}}(\theta) = \frac{\Gamma(N)}{\Gamma(M)}\frac{\Gamma(M+i\theta)}{\Gamma(N+i\theta)},
\end{equation}
implying that the PDF for the random variable $\log{p(\mathbf{m})}$ after $t$ time steps is given by
\begin{equation}
f_{\log{p(\mathbf{m})}}(x) = \left(\frac{\Gamma(N)}{\Gamma(M)}\right)^t\int_{-\infty}^{\infty} \frac{d\theta}{2\pi} e^{- i\theta x} \left(\frac{\Gamma(M+i\theta)}{\Gamma(N+i\theta)}\right)^t, \quad x < 0.
\end{equation}
Note that the constraint $x < 0$ closes the contour in the upper half-plane. This contour encloses all the poles of the integrand, since by the recurrence relation for Gamma functions the integrand has $N-M$ distinct poles of order $t$, 
\begin{equation}
\label{eq:gtheta}
g(\theta) = \left(\frac{\Gamma(M+i\theta)}{\Gamma(N+i\theta)}\right)^t =
i^{-(N-M)t}\prod_{j=0}^{N-M-1} \frac{1}{(\theta - i(M+j))^t},
\end{equation}
which are evenly spaced along the positive imaginary axis, $
\theta_j = i(M+j)$ for $j=0,1,\ldots,N-M-1$. Thus
\begin{equation}
f_{\log{p(\mathbf{m})}}(x) =  \left(\frac{\Gamma(N)}{\Gamma(M)}\right)^t 
\sum_{j=0}^{N-M-1} i \mathrm{Res}[e^{-i\theta x}g(\theta);\theta_j]
\end{equation}
with $g(\theta)$ given by Eq. \eqref{eq:gtheta}.

In principle, this calculation yields an exact expression for the p.d.f. $f_{\log{p(\mathbf{m})}}(x)$ of log Born probabilities for all times. In practice, these expressions rapidly become cumbersome. At $t=1$ we recover Eq. \eqref{eq:betadist} up to the necessary change of variables, while for $t=2$ we find that
\begin{equation}
f_{\log{p(\mathbf{m})}}(x) = \left(\frac{\Gamma(N)}{\Gamma(M)}\right)^2 e^{Mx}
\sum_{j=0}^{N-M-1} \frac{e^{jx}}{\Gamma(j+1)^2 \Gamma(N-M-j)^2}\left[2(\psi(j+1) - \psi(N-M-j)) -x\right].
\end{equation}
For all times, Eq. \eqref{eq:Borndist} implies that
\begin{equation}
\mathbb{E}[\log {p(\mathbf{m}})] = \mu t \sim \log{\left(\frac{1}{2^{pLt}}\right)}, \quad L \to \infty,
\end{equation}
in the large system limit, and similarly that
\begin{equation}
\mathrm{Var}[\log {p(\mathbf{m}})] = \varsigma^2 t \sim \frac{t}{\tau_P}, \quad L \to \infty.
\end{equation}
Thus we have proved that for large systems, the ``typical'' Born probability, i.e. the mean of $\log{p(\mathbf{m})}$, coincides with the uniform distribution over all $2^{pLt}$ possible measurement outcomes up to time $t$, while the distribution of $p(\mathbf{m})$ remains narrow on a log scale until times comparable to the purification time.

\subsection{Dynamics of R{\'e}nyi entropies}
We now illustrate how the above results provide an analytical handle on time evolution along quantum trajectories, starting from a maximally mixed initial state as in Eq. \eqref{eq:maxmixevo}. We will focus on the behaviour of R{\'e}nyi entropies with $\alpha >0$ (including the von Neumann entropy defined as the limit $\alpha \to 1$) along a given quantum trajectory, which can be expressed in terms of the corresponding Kraus operator's singular values as~\cite{nahum2021measurement}
\begin{equation}
\label{eq:RenyiSV}
S^{(\alpha)}_{\mathbf{m}}(t) = \frac{1}{1-\alpha} \log{\mathrm{Tr}[\rho^\alpha_{\mathbf{m}}(t)]} = \frac{1}{1-\alpha} \log{\sum_{n=1}^M\left(\frac{\sigma_n^{2}(t)}{\sum_{n'=1}^M\sigma_{n'}^{2}(t)}\right)^{\alpha}}.
\end{equation}
We discuss the limits of short and long times separately. Our analysis is similar in spirit to that of Ref. \cite{nahum2021measurement}, with a greater degree of analytical control owing to the results of previous sections.

\subsubsection{Short times}
\label{sec:shortproj}
We assume throughout this section that $L \gg 1$, so that the approximation of Kraus operators by a product $\hat{G}(t)$ of $t-1$ Ginibre matrices as in Eq. \eqref{eq:GinProd} holds, and let $\rho(\sigma,t) = \sum_{n=1}^M \delta(\sigma-\sigma_n(t))$ denote the density of singular values of $\hat{G}(t)$. Then the R{\'e}nyi entropies are determined by the ratios
\begin{equation}
\left(\frac{\sigma_n^{2}(t)}{\sum_{n'=1}^M\sigma_{n'}^{2}(t)}\right)^{\alpha} = \frac{\int_0^\infty d\sigma \, \rho(\sigma,t)\sigma^{2\alpha}}{\left(\int_0^\infty d\sigma \, \rho(\sigma,t)\sigma^2\right)^\alpha}.
\end{equation}
Let us write
\begin{equation}
\rho(\sigma,t) = \bar{\rho}(\sigma,t) + \delta \rho(\sigma,t),
\end{equation}
where $\bar{\rho}(\sigma,t) = \mathbb{E}[\rho(\sigma,t)]$, for the mean and fluctuations of $\rho(\sigma,t)$ about its mean respectively. Similarly, we write $m^{(\alpha)} = \int_0^\infty d\sigma \, \bar{\rho}(\sigma,t) \sigma^{2\alpha}$ and $\delta m^{(\alpha)} = \int_0^\infty d\sigma \, \delta \rho(\sigma,t) \sigma^{2\alpha}$. Then the ratio
\begin{equation}\label{eq:m}
\log{\frac{\int_0^\infty d\sigma \, \rho(\sigma,t)\sigma^{2\alpha}}{\left(\int_0^\infty d\sigma \, \rho(\sigma,t)\sigma^2\right)^\alpha}} = \log{\left(\frac{m^{(\alpha)}}{(m^{(1)})^{\alpha}}\right)} + \left(\frac{\delta m^{(\alpha)}}{m^{(\alpha)}} - \alpha \frac{\delta m^{(1)}}{m^{(1)}}\right) + \mathcal{O}(\delta \rho^2)
\end{equation}
is dominated by the moments of the mean spectral density, provided the leading fluctuation correction $\left(\frac{\delta m^{(\alpha)}}{m^{(\alpha)}} - \alpha \frac{\delta m^{(1)}}{m^{(1)}}\right)$ is small. Let us assume this at times that are short compared to the purification time (for an analytical argument that such short-time fluctuations are small in the context of weak measurements, see Section \ref{sec:shortweak}). Then the R{\'e}nyi entropies are dominated by their non-fluctuating part
\begin{equation}
\bar{S}^{(\alpha)}(t) = \frac{1}{1-\alpha} \log{\left(\frac{m^{(\alpha)}}{(m^{(1)})^{\alpha}}\right)}.
\end{equation}
In the regime of times $1 \ll t \ll M$, we have~\cite{Akemann_2019}
\begin{equation}
\label{eq:Akemannsemicircle}
\bar{\rho}(\sigma,t) \approx \begin{cases}
    \frac{2N^{1-\frac{1}{t}}\sigma^{\frac{2}{t}-1}}{t}, & \sigma^2 < M \left(\frac{M}{N}\right)^{t-1}, \\
    0, & \sigma^2 > M \left(\frac{M}{N}\right)^{t-1},
\end{cases}
\end{equation}
yielding
\begin{equation}
m^{(\alpha)} \approx \frac{1}{\alpha t + 1}\frac{M^{\alpha t+1}}{N^{\alpha(t-1)}},
\end{equation}
which predicts that
\begin{equation}
\label{eq:shorttimerenyi}
\bar{S}^{(\alpha)}(t) \approx \log M - \log t + \mathcal{O}(t^0), \quad 1 \ll t \ll M,
\end{equation}
for all $\alpha > 0$, recovering a result that had been previously derived in various specific limits using various distinct physical arguments~\cite{Gullans_2020,Li_2021,nahum2021measurement,ippoliti2023dynamical} and is consistent with our precise definition of the purification time in Eqs. \eqref{eq:purificationtime} and \eqref{eq:taupM}. 

For $\alpha = 2,3, \ldots,M$ (but not the von Neumann entropy) a more refined estimate is available, using the exact result~\cite{Akemann_2013}
\begin{equation}
m^{(\alpha)} = \frac{1}{\alpha !}\sum_{r=0}^{\alpha-1}(-1)^r \begin{pmatrix} \alpha-1 \\ r \end{pmatrix}  \left[\frac{(M-r-1+\alpha)!}{(M-r-1)!} \right]^{t}
\end{equation}
for such integer moments. This yields
\begin{equation}
\bar{S}^{(\alpha)}(t) = \frac{1}{1-\alpha} \log {\left(\frac{1}{\alpha! M^{\alpha t}} \sum_{r=0}^{\alpha-1}(-1)^r \begin{pmatrix} \alpha-1 \\ r \end{pmatrix} \left[\frac{(M-r-1+\alpha)!}{(M-r-1)!} \right]^{t}\right)}
\end{equation}
for the non-fluctuating contribution to the R{\'e}nyi entropies. The simplest non-trivial case of this formula is the second R{\'e}nyi entropy for which
\begin{equation}
\bar{S}^{(2)}(t) = - \log\left(\frac{1}{2}\left((1+1/M)^t - (1-1/M)^t\right)\right),
\end{equation}
which is easily verified to recover Eq. \eqref{eq:shorttimerenyi} in the limit $t \ll M$. 

\subsubsection{Long times}
\label{sec:longproj}
We next consider the dynamics of R{\'e}nyi entropies at times that are long compared to the purification time, for which the leading singular values of the Kraus operators are well separated from one another with high probability, so that the semicircle law Eq. \eqref{eq:Akemannsemicircle} no longer provides a good approximation to their distribution. First note that as $t \to \infty$ the squared singular values are distributed as~\cite{Forrester_2015,Akemann_2015}
\begin{equation}
\label{eq:ergodicity}
\sigma_n^2(t) \sim e^{2Y_nt},
\end{equation}
where the $Y_n$ are independent normal variables with mean $\lambda_n$ and variance
\begin{equation}
\varsigma_n^2 = \frac{1}{4t}(\psi'(M-n+1)-\psi'(N-n+1)).
\end{equation}
In particular, the R{\'e}nyi entropies and von Neumann entropies are dominated by the largest two singular values with high probability as $t \to \infty$, implying that
\begin{equation}
\label{eq:renyi}
S_{\mathbf{m}}^{(\alpha)}(t) \sim \begin{cases}
\frac{\alpha}{\alpha-1} \nu(t), & \alpha > 1, \\
\nu(t) \log{\nu(t)}, & \alpha = 1, \\
\frac{1}{1-\alpha} \nu(t)^{\alpha}, & 0<\alpha < 1,
\end{cases}
\end{equation}
in terms of the ratio $\nu(t)$ defined in Eq. \eqref{eq:ratio} (similar expressions were obtained in Appendix E of Ref. \cite{nahum2021measurement}). By our definition of the purification time Eq. \eqref{eq:purificationtime}, it follows that
\begin{equation}
\label{eq:logrenyi}
\mathbb{E}[\log{S_{\mathbf{m}}^{(\alpha)}(t)}] \sim \begin{cases}
-\frac{t}{\tau_\mathrm{P}}, & \alpha \geq 1, \\
-\frac{\alpha t}{\tau_\mathrm{P}}, & 0<\alpha < 1,
\end{cases}
\end{equation}
as $t \to \infty$. Thus we have constructed a solvable model that confirms earlier proposals for the late-time dynamics of entropy in entangling phases of spatially local monitored systems~\cite{Gullans_2020,Li_2021,nahum2021measurement}, together with the idea that at long times $t \gg \tau_P$, the purification time $\tau_{\mathrm{P}}$ captures the typical~\cite{nahum2021measurement} behaviour of a broad (and to a first approximation log-normal~\cite{Akemann_Lyap_2014,Akemann_2015}) distribution of $S_{\mathbf{m}}^{\alpha}(t)$.

Unfortunately, a more detailed characterization of the late-time distributions of $\nu(t)$ and $S_{\mathbf{m}}^{\alpha}(t)$, including their mean (rather than typical) values, is beyond the scope of the long-time prediction Eq. \eqref{eq:ergodicity} and related expressions~\cite{Akemann_Lyap_2014}, because these results do not contain information about the joint distribution of $\sigma_1^2(t)$ and $\sigma^2_2(t)$ that would be necessary to accurately model $\nu(t)$. We therefore turn to a model with weak measurements, for which the joint distribution function of singular values can be obtained analytically.

\section{Weak measurements}
\label{sec:secweak}
\subsection{Model and Fokker-Planck equation}
We now consider a generalization of Model II defined in Section \ref{sec:RankCollapseandModels} above, whose only distinction from Model II is that its measurement layers consist of independent weak measurements on $pL$ qubits per layer, which act on a single qubit as
\begin{equation}
\label{eq:1qubitweak}
\hat{P}_{\uparrow} = \frac{1}{2} \begin{pmatrix} 1+ \epsilon & 0 \\ 0 & 1-\epsilon \end{pmatrix}, \quad \hat{P}_{\downarrow} = \frac{1}{2} \begin{pmatrix} 1-\epsilon & 0 \\ 0 & 1+\epsilon \end{pmatrix},
\end{equation}
where we fix $0\leq \epsilon \leq 1$. We again write the Kraus operators along a single quantum trajectory at time $t$ as $
\hat{K}_{\mathbf{m}}(t) = \hat{P}_{\mathbf{m}_t} \hat{U}_t \ldots \hat{P}_{\mathbf{m}_1}\hat{U}_1$,
where $\hat{P}_{\mathbf{m}_j}$ now denotes the tensor product of $pL$ weak measurement operators as in Eq. \eqref{eq:1qubitweak}, and the identity on the remaining $(1-p)L$ qubits.  Regardless of the specific set of measurement outcomes, which we mostly suppress for economy of notation, we can write
\begin{equation}
\hat{P}_{\mathbf{m}_{j}} = \frac{1}{2^{pL}} (\hat{\mathbbm{1}}+\hat{\Lambda}_{j}),
\end{equation}
where $\hat{\Lambda}_{j}$ has eigenvalues
\begin{equation}
l_n = (1+\epsilon)^n(1-\epsilon)^{pL-n}-1, \quad n=0,1,\ldots,pL,
\end{equation}
with respective multiplicities
\begin{equation}
d_n = \begin{pmatrix} pL \\ n \end{pmatrix} 2^{(1-p)L}.
\end{equation}
It follows from these expressions that
\begin{equation}
\mathrm{tr}[\hat{\Lambda}_{j}] = 0, \quad \mathrm{tr}[\hat{\Lambda}_{j}^2] = 2^L\left[(1+\epsilon^2)^{pL}-1\right].
\end{equation}
We would like to understand the evolution of the singular values $\sigma_1(t) \geq \ldots \geq \sigma_N(t) \geq 0$ of $\hat{K}_{\mathbf{m}}(t)$. Thus consider the singular value decomposition $\hat{K}_{\mathbf{m}}(t) = \hat{V}_t\hat{D}_t \hat{W}^\dagger_t$ as in Eq. \eqref{eq:SVD}. Then
\begin{equation}
\hat{K}_{\mathbf{m}}(t+1)\hat{K}_{\mathbf{m}}(t+1)^\dagger = \hat{P}_{t+1} \hat{U}_{t+1} \hat{V}_{t} \hat{D}^2_t \hat{V}_t^\dagger \hat{U}_{t+1}^\dagger \hat{P}_{t+1}.
\end{equation}
Letting $\hat{U} = \hat{U}_{t+1} \hat{V}^\dagger_{t}$, which inherits Haar randomness of $\hat{U}_{t+1}$ at time $t+1$, we can write
\begin{equation}
\hat{U}^\dagger \hat{K}_{\mathbf{m}}(t+1) \hat{K}_{\mathbf{m}}(t+1)  \hat{U} = (\hat{U}^\dagger \hat{P}_{t+1} \hat{U})\hat{D}^2_t (\hat{U}^\dagger \hat{P}_{t+1} \hat{U}).
\end{equation}
Introducing the operators $\hat{B} = \hat{U}^\dagger \hat{\Lambda}_{t+1} \hat{U}$ and $\hat{X}_t = 2^{2pLt}\hat{K}_{\mathbf{m}}(t)\hat{K}_{\mathbf{m}}(t)^\dagger$, we have
\begin{equation}
\label{eq:recurrence}
\hat{X}_{t+1} \overset{\mathrm{s.v.}}{\sim} (\hat{\mathbbm{1}}+\hat{B}) \hat{X}_t (\hat{\mathbbm{1}}+\hat{B}) = \hat{X}_t + \hat{B}\hat{X}_t + \hat{X}_t\hat{B} + \hat{B} \hat{X}_t \hat{B}.
\end{equation}
where as above $\overset{\mathrm{s.v.}}{\sim}$ denotes a common distribution of singular values (which are eigenvalues in this case). Let $x_1(t) \geq \ldots \geq x_N(t) \geq 0$ denote the eigenvalues of $\hat{X}_t$, which are given by $x_n(t) = 2^{2pLt}\sigma_n^2(t)$ in terms of the singular values $\sigma_n(t)$ of the Kraus operators. Then, if $pL\epsilon^2 \ll 1$ so that typical eigenvalues of $\Lambda$ are small assuming large $pL$, Eq. \eqref{eq:recurrence} can be expanded perturbatively in $\epsilon$ to yield
\begin{equation}
x_{n}(t+1) = x_n(t) + B_{nn} x_n(t) + x_n(t) B_{nn} + \sum_{m} |B_{nm}|^2 x_m(t) + \sum_{m \neq n} |B_{nm}|^2 \frac{(x_n(t)+x_m(t))^2}{x_n(t)-x_m(t)} + \mathcal{O}(\epsilon^3)
\end{equation}
Collecting terms yields a perturbative model for the time evolution of the singular values $x_n(t)$, namely
\begin{equation}
x_n(t+1)-x_n(t) = \left(2B_{nn}+ \sum_{m} |B_{nm}|^2 +  4 \sum_{m \neq n} |B_{nm}|^2 \frac{x_m(t)}{x_n(t) - x_m(t)}\right)x_n(t).
\end{equation}
This equation splits naturally into multiplicative ``noise'' and ``drift'' contributions, given by
\begin{equation}
\eta_n(t+1) = 2 B_{nn}, \quad \Delta_n(t+1) = \sum_{m} |B_{nm}|^2 +  4 \sum_{m \neq n} |B_{nm}|^2 \frac{x_m(t)}{x_n(t) - x_m(t)}.
\end{equation}
Thus
\begin{equation}
x_n(t+1)-x_n(t) = \left(\eta_n(t+1) + \Delta_n(t+1)\right) x_n(t).
\end{equation}
To eliminate multiplicative noise, we write $x_n(t) = e^{y_n(t)}$ and, again assuming that $\epsilon$ is small, find that 
\begin{equation}
y_n(t+1)-y_n(t) = \log{\left(\eta_n(t+1) + \Delta_n(t+1)\right)} = \eta_n(t+1) + \mathcal{D}_n(t+1)+ \mathcal{O}(\epsilon^3),
\end{equation}
where the new drift term
\begin{equation}
\mathcal{D}_n(t+1) = \Delta_n(t+1) - \frac{1}{2}\eta_n(t+1)^2.
\end{equation}

To proceed further, we note that the matrix elements
\begin{equation}
B_{nn} =\sum_{a} U_{an}U^*_{an}(\Lambda(t+1))_{aa}
\end{equation}
and
\begin{equation}
|B_{mn}|^2 = \sum_{a,b} U_{an}U_{bm}U^*_{am}U^*_{bn} (\Lambda(t+1))_{aa}(\Lambda(t+1))_{bb}
\end{equation}
are amenable to Haar averaging. In particular, the two-point function
\begin{equation}
\langle U_{an} U^*_{an} \rangle = \frac{1}{N},
\end{equation}
and the four-point functions
\begin{align}
 \langle U_{an} U_{bm} U^*_{an} U^*_{bm} \rangle &= \frac{1}{N^2-1}(1+\delta_{ab}\delta_{mn}) - \frac{1}{N(N^2-1)}(\delta_{mn}+\delta_{ab}), \\
\langle U_{an}U_{bm}U^*_{am}U^*_{bn}  \rangle &= \frac{1}{N^2-1}(\delta_{mn}+\delta_{ab}) -\frac{1}{N(N^2-1)} (1+\delta_{ab}\delta_{mn}),
\end{align}
by standard results~\cite{creutz1978invariant}, where angle brackets $\langle...\rangle$ denote the Haar average over $\hat{U} \in SU(N)$, implying that
\begin{equation}
\langle \eta_n(t) \rangle = \frac{1}{N}\mathrm{tr}[\hat{\Lambda}] =0
\end{equation}
and
\begin{equation}
\langle \eta_m(t_1)\eta_n(t_2) \rangle = \Gamma \delta_{t_1t_2} \delta_{mn}
\end{equation}
where the noise strength
\begin{equation}
\Gamma = \frac{4}{N^2-1}\left(1-\frac{1}{N}\right) \mathrm{tr}[\hat{\Lambda}^2] \approx \frac{4}{N^2} \mathrm{tr}[\hat{\Lambda}^2]
\end{equation}
for $L \gg 1$. Similarly $
\langle |B_{mn}|^2 \rangle = \frac{1}{N^2-1} \left(1-\frac{\delta_{mn}}{N}\right)\mathrm{tr}[\hat{\Lambda}^2] \approx \frac{\Gamma}{4}$, implying that the average value of the drift term (conditioned on the circuit realization at time $t$) is given by
\begin{equation}
\langle \mathcal{D}_n(t+1) \rangle \approx \frac{\Gamma(N-2)}{4} + \Gamma \sum_{m\neq n} \frac{e^{y_m(t)}}{e^{y_n(t)}- e^{y_m(t)}}
\end{equation}
Thus, assuming that fluctuations of $\mathcal{D}_{n}(t+1)$ about its Haar-averaged value are negligible, we obtain the Langevin equation
\begin{equation}
y_{n}(t+1) - y_n(t) = \eta_n(t+1) + \frac{\Gamma(N-2)}{4} + \Gamma \sum_{m\neq n} \frac{e^{y_m(t)}}{e^{y_n(t)}-e^{y_m(t)}}.
\end{equation}
This can be written in a more symmetric form by making the change of variables $2z_n(t) = y_n(t) + \frac{\Gamma N}{4}t$, which yields
\begin{equation}
z_n(t+1)-z_n(t) = \frac{1}{2}\eta_n(t) + \frac{\Gamma}{4} \sum_{m\neq n} \coth{(z_n(t)-z_m(t))}.
\end{equation}
Finally, letting $\Gamma \to 0$ yields the continuous-time Fokker-Planck equation
\begin{align}
\label{eq:FP}
\partial_s P  = \sum_{n=1}^N\left(-\partial_{z_n}(D_n P)  + \partial_{z_n}^2 P\right),
\end{align}
where $P(\vec{z},s)$ denotes the joint PDF of $\vec{z}(t) = (z_1(t),z_2(t),\ldots,z_N(t))$ at time 
\begin{equation}
\label{eq:defofs}
s = \Gamma t/8,
\end{equation} 
and
\begin{equation}
D_n(\vec{z}) = 2 \sum_{m \neq n} \coth{(z_n-z_m)}.
\end{equation}
In particular, the timescale $\Gamma^{-1}$ determines the dynamics of singular values, and should be thought of as the purification time for these models, an interpretation that will be confirmed by our results for the short and long time dynamics of R{\'e}nyi entropies below.

The continuous-time Fokker-Planck equation Eq. \eqref{eq:FP} was first derived~\cite{Ipsen_2016} in the context of a stochastic matrix model known as ``isotropic Brownian motion''~\cite{le1985isotropic}. In that work, it was also pointed out that Eq. \eqref{eq:FP} is exactly solvable via a connection to Calogero-Sutherland models. In order to be self-contained, we now derive this exact solution.

\subsection{Exact solution of the Fokker-Planck equation}

To solve Eq. \eqref{eq:FP}, we first note that $D_n$ can be written in gradient form
\begin{equation}
D_n(\vec{z}) = -\partial_{z_n}\Phi(\vec{z})
\end{equation}
where the ``prepotential''
\begin{equation}
\Phi(\vec{z}) = -2 \sum_{j<k} \log{\sinh{(z_j-z_k)}}
\end{equation}
arises naturally in the theory of the classical hyperbolic Calogero-Sutherland model~\cite{Kulkarni_2017}. This implies that Eq. \eqref{eq:FP} can be written as~\cite{risken1996fokker}
\begin{equation}
\partial_s P = \partial_{z_n} \left(e^{-\Phi} \partial_{z_n} \left(e^{\Phi}P\right)\right).
\end{equation}
Letting $\psi = e^{\Phi/2} P$ then yields an imaginary-time Schr{\"o}dinger equation for $\psi$, namely~\cite{sutherland1972exact,risken1996fokker}
\begin{equation}
\partial_s \psi  = \sum_{n=1}^N \partial_{z_n}^2 \psi - V \psi, 
\end{equation}
with an effective potential
\begin{equation}
\label{eq:Veff}
V(\vec{z}) = \frac{1}{4} \sum_{n=1}^N D_n^2(\vec{z}) + \frac{1}{2} \sum_{n=1}^N \partial_{z_n}D_n(\vec{z}).
\end{equation}
To proceed further write $z_{nm} = z_n-z_m$ and note that
\begin{equation}
 \frac{1}{4} \sum_{n=1}^N D_n^2(\vec{z}) = \sum_{\substack{n \neq m \\n \neq l}} \coth{z_{nm}}\coth{z_{nl}}  = \sum_{n \neq m} \coth^2{z_{nm}} + \sum_{\substack{l \neq m \\ m \neq n \\ n \neq l}} \coth{z_{nm}}\coth{z_{nl}}.
\end{equation}
By an identity apparently first published by Calogero and Perelomov~\cite{calogero1978properties}, we have
\begin{equation}
\sum_{\substack{l \neq m \\ m \neq n \\ n \neq l}} \coth{z_{nm}}\coth{z_{nl}} = \frac{1}{3}N(N-1)(N-2)
\end{equation}
while
\begin{equation}
\frac{1}{2} \sum_{n=1}^N \partial_{z_n}D_n(\vec{z}) = N(N-1) - \sum_{n \neq m} \coth^2{z_{nm}},
\end{equation}
yielding a complete cancellation of interactions in the effective potential
\begin{equation}
V(\vec{z}) = \frac{1}{3}N(N^2-1),
\end{equation}
which is reminiscent of (but simpler than) the cancellation of interactions that occurs for the $\beta=2$ case of the DMPK equation~\cite{beenakker1993nonlogarithmic}. Making the change of variables $\psi = e^{-\frac{N(N^2-1)}{3}s}\tilde{\psi}$ finally reduces the Fokker-Planck evolution Eq. \eqref{eq:FP} to a free diffusion equation
\begin{equation}
\label{eq:nonintdiff}
\partial_s \tilde{\psi}  = \sum_{n=1}^N \partial_{z_n^2} \tilde{\psi},
\end{equation}
(albeit with non-trivial boundary conditions to be discussed shortly). 

It remains to solve Eq. \eqref{eq:nonintdiff} in the ``ordered sector'' $z_1 \geq z_2 \geq \ldots \geq z_N$, subject to the initial condition
\begin{equation}
\label{eq:initcond} P(\vec{z},0) = \prod_{j=1}^N \delta(z_j-\varepsilon_j),
\end{equation}
where the regulators $\varepsilon_1 > \varepsilon_2 > \ldots > \varepsilon_N > 0$ will be taken to zero at the end of the calculation and are needed to avoid the singularities in Eq. \eqref{eq:FP} at collision planes $z_j = z_k$. We further impose boundary conditions
\begin{equation}
\label{eq:boundarycond}
e^{-\Phi(\vec{z})} \left(\partial_{z_j}-\partial_{z_k}\right)\left(e^{\Phi(\vec{z})} P(\vec{z},s) \right) = 0, \quad z_j \to z_k^+, \quad j < k,
\end{equation}
for all $s \geq 0$ that are sufficient for the vanishing of probability flux at collision planes, and therefore guarantee conservation of probability within the ordered sector.

Our method of solution follows that of Refs. \cite{beenakker1993nonlogarithmic,beenakker1994exact} for the $\beta=2$ DMPK equation but differs in its details. It is first useful to note that for every solution to Eq. \eqref{eq:nonintdiff}, there exists a solution to \eqref{eq:FP}, given by
\begin{equation}
\label{eq:defofwf}
P(\vec{z},s) = e^{-\frac{N(N^2-1)}{3}s} \left(\frac{\prod_{j<k} \sinh(z_j-z_k)}{\prod_{j<k} \sinh(\varepsilon_j-\varepsilon_k)}\right)\tilde{\psi}(\vec{z},s).
\end{equation}
In terms of the ``wavefunction'' $\tilde{\psi}$ in Eq. \eqref{eq:defofwf}, we find that the initial condition
\begin{equation}
\tilde{\psi}(\vec{z},0) = \prod_{j=1}^N \delta(z_j-\varepsilon_j)
\end{equation}
and the vanishing of the wavefunction at collision planes,
\begin{equation}
\tilde{\psi}(\vec{z},s) = 0, \quad z_j \to z_k^+, \quad j < k,
\end{equation}
for $s \geq 0$, are sufficient for the initial and boundary conditions Eqs. \eqref{eq:initcond} and  
Eq. \eqref{eq:boundarycond} on $P$ to hold. 
To satisfy the boundary conditions in the ordered sector, we extend $\tilde{\psi}$ to the whole space and consider a ``fermionic'' initial condition
\begin{equation}
\tilde{\psi}(\vec{z},0) = \sum_{\kappa \in S_N} (-1)^{\mathrm{sgn}(\kappa)} \prod_{j=1}^N \delta(z_j-\varepsilon_{\kappa(j)}), 
\end{equation}
where $\mathrm{sgn}: S_N \to {\pm 1}$ denotes the sign of the permutation $\kappa \in S_N$. We can write this more suggestively as a Slater determinant of single-particle factors
\begin{equation}
\tilde{\psi}(\vec{z},0) = \mathrm{det}\left[\delta(z_j-\varepsilon_k)\right],
\end{equation}
where $\mathrm{det}[A_{jk}]$ denotes the determinant of the matrix $A$ with elements $A_{jk}$. Then the Slater determinant of heat kernels
\begin{equation}
\label{eq:heatdet}
\tilde{\psi}(\vec{z},s) = \mathrm{det}\left[\frac{1}{\sqrt{4 \pi s}}e^{-\frac{(z_j-\varepsilon_k)^2}{4s}}\right]
\end{equation}
satisfies both the diffusion equation Eq. \eqref{eq:nonintdiff} for $s \geq 0$ and the initial and boundary conditions Eqs. \eqref{eq:initcond} and \eqref{eq:boundarycond} in the ordered sector. The final nontrivial step is to set the regulators $\varepsilon_j \to 0$. 

To this end, we first note that the leading behaviour of the denominator of Eq. \eqref{eq:defofwf}
\begin{equation}
\label{eq:leadingeps}
\prod_{j<k} \sinh{(\varepsilon_j - \varepsilon_k)} \sim \prod_{j<k} (\varepsilon_j - \varepsilon_k), \quad \varepsilon_1 \ll 1,
\end{equation}
is a Vandermonde determinant of order $N-1$ in each of the $\varepsilon_j$. We next expand the heat kernels in terms of Hermite polynomials~\cite{abramowitz1968handbook}
\begin{equation}
\label{eq:HeatKernel}
e^{-\frac{(z_j-\varepsilon_k)^2}{4s}} 
\sim e^{-z_j^2/4s}  \sum_{n=0}^{N-1} H_n(\tilde{z}_j)\frac{\tilde{\varepsilon}_k^n}{n!}, \quad \varepsilon_1 \ll 1,
\end{equation}
to the same order in $\varepsilon$, where $\tilde{z}_j = z_j/\sqrt{4s}, \, \tilde{\varepsilon}_k = \varepsilon_k/\sqrt{4s}$. However, at this order, the sum in Eq. \eqref{eq:HeatKernel} is nothing but a product of square matrices and so
\begin{align}
\nonumber 
\mathrm{det}\left[e^{-z_j^2/4s}  \sum_{n=0}^{N-1} H_n(\tilde{z}_j)\frac{\tilde{\varepsilon}_k^n}{n!}\right] &= \frac{e^{-|\vec{z}|^2/4s}}{\prod_{n=1}^{N-1}n!} 
\begin{vmatrix} H_0(\tilde{z}_1) & H_1(\tilde{z}_1) & \ldots & H_{N-1}(\tilde{z}_1) \\
H_0(\tilde{z}_2) & H_1(\tilde{z}_2) & \ldots & H_{N-1}(\tilde{z}_2) \\
\vdots & \vdots & \ddots & \vdots \\ 
H_0(\tilde{z}_N) & H_1(\tilde{z}_N) & \ldots & H_{N-1}(\tilde{z}_N) \\
\end{vmatrix}
\begin{vmatrix} 1 & 1 & \ldots & 1 \\
\tilde{\varepsilon}_1 & \tilde{\varepsilon}_2 & \ldots & \tilde{\varepsilon}_N \\
\vdots & \vdots & \ddots & \vdots \\ 
\tilde{\varepsilon}_1^{N-1} & \tilde{\varepsilon}_2^{N-1} & \ldots & \tilde{\varepsilon}_N^{N-1}
\end{vmatrix}
\\
= &\frac{1}{(2s)^{N(N-1)/2}\prod_{n=1}^{N-1}n!} e^{-|\vec{z}|^2/4s} \prod_{j<k} (\varepsilon_j - \varepsilon_k)(z_j - z_k),
\end{align}
where in the second line we applied Gaussian elimination to the matrix of Hermite polynomials to obtain another Vandermonde determinant. Combining the above expressions, we deduce that
\begin{equation}
\label{eq:exactpdf}
P(\vec{z},s) = \frac{1}{(4\pi s)^{N/2}(2s)^{N(N-1)/2}\prod_{n=1}^{N-1}n!}  e^{-\frac{N(N^2-1)}{3}s} \left(\prod_{j<k}(z_j-z_k)\sinh(z_j-z_k) \right)e^{-|\vec{z}|^2/4s}
\end{equation}
solves the Fokker-Planck equation for all $s \geq 0$, satisfies the initial condition $
P(\vec{z},0) = \prod_{j=1}^N \delta(z_j)$, and conserves probability in the ordered sector. This recovers the solution to Eq. \eqref{eq:FP} presented in Ref. \cite{Ipsen_2016}. We note for future reference that the random variables $z_n(t)$ modelled by Eq. \eqref{eq:exactpdf} are related to the singular values $\sigma_n(t)$ of Kraus operators at time $t$ by the change of variables 
\begin{equation}
\label{eq:zinsigma}
e^{z_n(t)} = e^{\left(pL\log{2} + \frac{\Gamma N}{8}\right)t} \sigma_n(t)
\end{equation}
and recall that $s$ is related to $t$ by Eq. \eqref{eq:defofs}.

\subsection{Very short times: emergence of log-GUE}
\label{sec:veryshort}
Let us first consider the probability distribution Eq. \eqref{eq:exactpdf} as $s \to 0$ (note that $s \ll 1$ corresponds to $t \ll \Gamma^{-1}$, i.e. times that are short compared to the purification time). We have
\begin{equation}
P(\vec{z},s) = \frac{1}{(4\pi s)^{N/2}(2s)^{N(N-1)/2}\prod_{n=1}^{N-1}n!} \left(\prod_{j<k}(z_j-z_k)^2 + \mathcal{O}(s^2)\right)e^{-|\vec{z}|^2/4s}.
\end{equation}
Thus the leading contribution to $P(\vec{z},s)$ is from the distribution function $P_{\mathrm{v.s.t.}}(\vec{z},s)$, given by
\begin{equation}
\label{eq:maptoGUE}
P(\vec{z},s) \sim P_{\mathrm{v.s.t.}}(\vec{z},s) = \frac{(2s)^{N(N-1)/2}}{(4\pi s)^{N/2}\prod_{n=1}^{N-1}n!} \left(\prod_{j<k}\frac{(z_j-z_k)^2}{4s}\right)e^{-|\vec{z}|^2/4s}, \quad s \to 0^+.
\end{equation}
This is strongly reminiscient of the distribution function of the Gaussian Unitary Ensemble (GUE) on the full space~\cite{mehta2004random},
\begin{equation}
P_{\mathrm{GUE}}(\vec{z}) = \frac{1}{N!} \frac{2^{N(N-1)/2}}{\pi^{N/2} \prod_{n=1}^{N-1}n!} \left(\prod_{j<k}(z_j-z_k)^2\right)e^{-|\vec{z}|^2}
\end{equation}
and indeed the two distributions are related by a simple change of variables
\begin{equation}\label{vst}
P_{\mathrm{v.s.t.}}(\vec{z},s) = \frac{N!}{(4s)^{N/2}} P_{\mathrm{GUE}}(\vec{z}/\sqrt{4s}),
\end{equation}
where the prefactor of $N!$ arises because we restricted the domain of $P(\vec{z},s)$ to the ordered sector. We will call the regime of validity of the approximation Eq. \eqref{eq:maptoGUE} the ``very-short-time regime'', to be determined below. In the very-short-time regime, the singular values of Kraus operators follow a ``log-GUE'' distribution, which is related to the conventional GUE in the same sense that the log-normal distribution is related to the normal distribution. Defining the random variable $\rho(z,t) = \sum_{n=1}^N \delta(z-z_n(t))$, it follows by the Wigner semicircle law for the conventional GUE~\cite{mehta2004random} that the non-fluctuating part of the level density
\begin{equation}
\label{eq:vstsemicircle}
\bar{\rho}_{\mathrm{v.s.t.}}(z,t) = \begin{cases} \frac{2}{\pi} \frac{1}{\Gamma t} \sqrt{ N \Gamma t - z^2}, & |z| \leq \sqrt{N \Gamma t}, \\ 0, &  |z| > \sqrt{N \Gamma t}, \end{cases}
\end{equation}
for $N \gg 1$ and very short times. The boundary of the very-short-time regime is set by the requirement in Eq.~\eqref{vst} that $|z_j-z_k| \ll 1$ for all $j,k$: from Eq.~\eqref{eq:vstsemicircle}, this is the case for $N\Gamma t \ll 1$.

We can use this observation to extend the analysis of R{\'e}nyi entropies in Section \ref{sec:shortproj} to very short times as follows. First note that in terms of $\rho(z,t)$, the R{\'e}nyi entropies along a given quantum trajectory can be written as
\begin{equation}
S_{\mathbf{m}}^{(\alpha)}(t) = \frac{1}{1-\alpha} \log{\left(\frac{\int_{-\infty}^\infty dz \, \rho(z,t) e^{2\alpha z}}{\left(\int_{-\infty}^\infty dz \, \rho(z,t)e^{2z}\right)^\alpha}\right)}.
\end{equation}
In particular, we have
\begin{equation}
\int_{-\infty}^{\infty} dz \, \bar{\rho}_{\mathrm{v.s.t.}}(z,t) e^{2\alpha z} = \frac{2N}{\pi} \int_0^\pi d\theta \, \sin^2{\theta} e^{2\alpha \sqrt{N \Gamma t} \cos{\theta}} = \frac{1}{\alpha} \sqrt{\frac{N}{\Gamma t}} I_1(2\alpha \sqrt{N \Gamma t}) ,
\end{equation}
where $I_1(w)$ denotes a modified Bessel function of the first kind~\cite{abramowitz1968handbook}, so that the non-fluctuating part of each R{\'e}nyi entropy is given by
\begin{equation}
\bar{S}^{(\alpha)}(t) = \frac{1}{1-\alpha} \log{ \left(\frac{1}{\alpha} \left(\frac{N}{\Gamma t}\right)^{\frac{1-\alpha}{2}} \frac{I_1(2\alpha \sqrt{N \Gamma t})}{I_1(2\sqrt{N \Gamma t})^\alpha}\right)}, \quad \Gamma t \ll 1.
\end{equation}
The small-argument asymptotic behaviour
\begin{equation}
I_1(w) \sim \frac{w}{2}(1+\frac{w^2}{8}), \quad w \to 0,
\end{equation}
implies perturbative time dependence
\begin{equation}
\bar{S}^{(\alpha)}(t) \approx \log{N} - \frac{\alpha N \Gamma t}{2}, \quad \Gamma t \ll \frac{1}{N},
\end{equation}
of the R{\'e}nyi entropies at very short times. Meanwhile, log-GUE statistics and the semicircle law Eq. \eqref{eq:vstsemicircle} break down at times $\frac{1}{N} \ll \Gamma t \ll 1$ that are short but not very short, as we now discuss.

\subsection{Short times: semicircle-to-square crossover}
\label{sec:shortweak}
Let us now consider how the semicircle law corresponding to the exact distribution function $P(\vec{z},s)$ differs from Eq. \eqref{eq:vstsemicircle} beyond the regime of very short times. We first note that $P(\vec{z},s)$ can be written as a Boltzmann weight for a harmonically trapped gas of particles on a line at positions $z_1 <z_2 < \ldots <z_N$ in the standard fashion,
\begin{equation}
P(\vec{z},s) \propto e^{-W(\vec{z},s)},
\end{equation}
where the effective potential
\begin{equation}
W(\vec{z},s) = \sum_{i=1}^N \frac{z_i^2}{4s} -\frac{1}{2}\sum_{j \neq k} \log{\left((z_j-z_k)\sinh{(z_j-z_k)}\right)}
\end{equation}
interpolates between Dyson's ``log-gas''~\cite{mehta2004random} in the limit of small interparticle separations~\cite{Ipsen_2016} $|z_j-z_k| \to 0$ that was discussed in the previous section, and a one-dimensional Coulomb gas at large interparticle separations~\cite{forrester2022global} $|z_j-z_k| \to \infty$, as follows from the asymptotic behaviour $\log{(z\sinh{z})} \sim \log{(|z|)}$ as $|z| \to \infty$. The latter model is also known in the literature as the ``one-dimensional jellium model'', various properties of which can be derived analytically~\cite{baxter1963statistical,flack2022gap,flack2023exact}, such as its uniform density of states~\cite{baxter1963statistical}. 

It will be instructive to understand this crossover to uniformity in terms of the density of states $\rho(z)$. Suppressing explicit time dependence, we have
\begin{equation}
\label{eq:energyfunctional}
W[\rho] = \int_{-\infty}^\infty dz \, \rho(z) \frac{z^2}{4s} - \frac{1}{2}\int_{-\infty}^\infty \int_{-\infty}^\infty dz \, dw \, \rho(z) \rho(w) \log{\left((z-w)\sinh{(z-w)}\right)}.
\end{equation}
Note also that $\rho(z)$ satisfies the constraint $N[\rho] = \int_{-\infty}^{\infty} dz \, \rho(z) = N$. Thus for large $N \gg 1$, the non-fluctuating part of the density of states $\bar{\rho}(z)$ minimizes the functional $W[\bar{\rho}] - \mu N[\bar{\rho}]$, implying that
\begin{equation}
\frac{z^2}{4s} - \int_{-\infty}^{\infty} dw \, \bar{\rho}(w) \log{\left((z-w)\sinh{(z-w)}\right)} = \mu.
\end{equation}
Finally differentiating with respect to $z$ yields the singular integral equation~\cite{muskhelishvili1953singular,mergny2021stability}
\begin{equation}
\label{eq:sie}
\dashint_{-\infty}^{\infty} dw \, \bar{\rho}(w) \left(\frac{1}{z-w} + \coth{(z-w)}\right) = \frac{z}{2s},
\end{equation}
which is distinguished from the usual semicircle law by the presence of the $\coth{(z-w)}$ term in the kernel, that breaks the spatial rescaling symmetry of the former. This breaking of scaling symmetry allows $\bar{\rho}$ to flow from a semicircle law to a uniform distribution as $s$ increases. We note that an exact but implicit solution to the integral equation Eq. \eqref{eq:sie} was proposed recently in the literature~\cite{mergny2021stability}.

\begin{figure}[t]
    \centering
    \includegraphics[width= 0.75\linewidth]{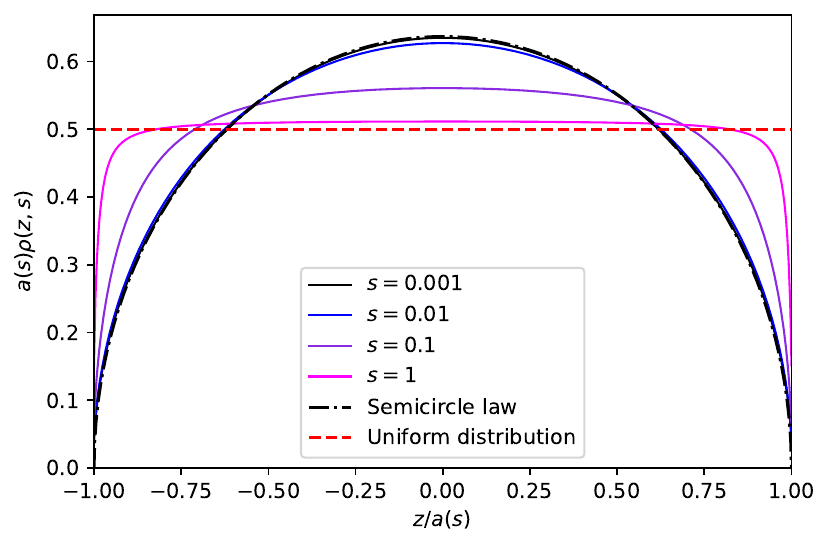}
    \caption{Numerical solutions to the singular integral equation Eq. \eqref{eq:sie} on intervals $[-a(s),a(s)]$ with the endpoint position $a(s)$ determined implicitly by number conservation. We set $N=50$ and discretize the integral over one thousand points. The accuracy of our scheme is confirmed by its recovery of the semicircle law discussed in the previous section as $s \to 0$, and we observe a clear semicircle-to-square crossover from a semicircle law to uniform behaviour between the very-short-time ($s \ll 0.01$) and short-time ($0.01 \ll s \ll 1$) regimes defined by Eq. \eqref{eq:endpoint}.}
    \label{FigSemicirc}
\end{figure}

We instead proceed numerically and solve for solutions to Eq. \eqref{eq:sie} supported on a finite interval $[-a,a]$ with $a>0$ for $s>0$ (note that the microscopic distribution function Eq. \eqref{eq:exactpdf} is inversion symmetric in $z$). Numerical results obtained by discretizing the principal value integral in Eq. \eqref{eq:sie} and inverting the resulting matrix to obtain $\rho(z,s)$ are shown in Fig. \ref{FigSemicirc} and confirm that the semicircle law discussed in the previous section quickly converges to a uniform distribution beyond the regime of very short times. The resulting late-time profile is sometimes called a ``square law''~\cite{Ipsen_2016} and in this sense Eq. \eqref{eq:sie} captures a ``semicircle-to-square'' crossover with increasing $s$.

Let us now attempt to understand this crossover to uniformity analytically. Thus consider the uniform ansatz 
\begin{equation}
\label{eq:uniformansatz}
\bar{\rho}_{\mathrm{s.t.}}(w) = \begin{cases} \frac{N}{2a},  & |w| \leq a, \\ 0, & |w| > a.
\end{cases}
\end{equation}
This yields
\begin{equation}
\label{eq:testingansatz}
\dashint_{-a}^{a} dw \, \bar{\rho}_{\mathrm{s.t.}}(w) \left(\frac{1}{z-w} + \coth{(z-w)}\right) = \frac{N}{2a} \log{\left(\frac{(a+z)\sinh{(a+z)}}{(a-z)\sinh{(a-z)}}\right)}.
\end{equation}
Expanding the right-hand side of Eq. \eqref{eq:testingansatz} perturbatively in $z$ yields a linear estimate $Nz/a$. We will show that Eq. \eqref{eq:testingansatz} is close to this linear estimate over a large ``bulk region'' $|z \pm a| > \varepsilon$, where $\varepsilon > 0$ is an order one constant. To this end, we define the function
\begin{equation}
\eta(z) = \frac{N}{2a}\left(\log{\left(\frac{(a+z)\sinh{(a+z)}}{(a-z)\sinh{(a-z)}}\right)} - 2z\right),
\end{equation}
which quantifies the deviation of the right-hand side of Eq. \eqref{eq:testingansatz} from linearity. The function $\eta(z)$ is odd and strictly increasing in $z$ on the interval $(-a,a)$. It is also convex on the interval $[0,a-\epsilon]$. Together these observations imply an upper bound
\begin{equation}
\frac{|\eta(z)|}{N|z|/a} \leq \frac{a}{N}\frac{\eta(a-\varepsilon)}{a-\varepsilon}, \quad 0 < |z| < a-\varepsilon,
\end{equation}
valid over the bulk region. Finally, we note that to leading order in $a$ and for $\varepsilon$ of order one,
\begin{equation}
\frac{a}{N}\frac{\eta(a-\varepsilon)}{a-\varepsilon} = \frac{\log{a}}{2a} + \mathcal{O}\left(\frac{1}{a}\right),
\end{equation}
which establishes that 
\begin{equation}
\dashint_{-a}^{a} dw \, \bar{\rho}_{\mathrm{s.t.}}(w) \left(\frac{1}{z-w} + \coth{(z-w)}\right) = \frac{Nz}{a}\left(1+ \mathcal{O}\left(\frac{\log{a}}{a}\right)\right)
\end{equation}
in the entire bulk region. Comparison with the original integral equation \eqref{eq:sie} reveals that the uniform ansatz Eq. \eqref{eq:uniformansatz} yields an accurate bulk solution provided we make the identification 
\begin{equation}
\label{eq:endpoint}
a(s) = 2sN \gg 1,
\end{equation}
corresponding to the regime of times $\Gamma t \gg \frac{1}{N}$, which strictly excludes the very-short-time regime identified in the previous section. Thus we have shown that from the short-time regime onwards, the uniform ansatz Eq. \eqref{eq:uniformansatz} yields a good approximation to the solution of Eq. \eqref{eq:sie}, excluding a boundary region of width $2\varepsilon$ that has negligible measure compared to the bulk region. The latter point in particular means that the uniform ansatz Eq. \eqref{eq:uniformansatz} can reliably be used to estimate integrals over singular values when the separation between neighbouring singular values is small. 

For example, at short times the uniform ansatz Eqs. \eqref{eq:uniformansatz} and \eqref{eq:endpoint} predicts that
\begin{equation}
\label{eq:weakshortrenyi}
\bar{S}^{(\alpha)}(t) = \frac{1}{1-\alpha} \log{\left(\frac{1}{\alpha} \left(\frac{2}{\Gamma t}\right)^{1-\alpha} \frac{\sinh(\alpha N \Gamma t /2)}{\sinh(N \Gamma t / 2)^\alpha}\right)} \approx - \log{\Gamma t}, \quad \frac{1}{N} \ll \Gamma t \ll 1,
\end{equation}
whose leading time dependence perfectly matches Eq. \eqref{eq:shorttimerenyi} for projective measurements, provided that $\Gamma^{-1}$ is interpreted as the purification time. That $\Gamma^{-1}$ indeed defines the purification time in the sense of Eq. \eqref{eq:purificationtime} will be checked carefully below, but can be seen here by noting that the timescale for the characteristic spacing between adjacent $z_n$ to become comparable to their long-time standard deviation, invalidating the treatment of $\rho(z,t)$ as a uniform distribution, is defined by the condition $2a(s)/N = \mathcal{O}(s^{1/2})$, i.e. $\Gamma t = \mathcal{O}(1)$.

Finally, we note that the energy functional Eq. \eqref{eq:energyfunctional} appearing in the Dyson gas formulation of our problem can be used to estimate the strength of fluctuations of the R{\'e}nyi entropies, thereby validating our estimation of short-time R{\'e}nyi entropies by their non-fluctuating values as in Eq. \eqref{eq:weakshortrenyi}. We first expand the functional $W$ perturbatively about the mean field $\bar{\rho}(w)$ to yield a quadratic effective action
\begin{equation}
\mathcal{S}[\delta \rho] = W[\bar{\rho}+\delta \rho] - W[\bar{\rho}] = \frac{1}{2}\int_{-a}^{a}\int_{-a}^{a} dz\, dw \, K(z-w) \delta \rho(z) \delta \rho(w),
\end{equation}
whose kernel
\begin{equation}
K(z-w) = -\log{(z-w)\sinh{(z-w)}}.
\end{equation}
It follows that the expected value of the correlation function $\delta \rho(z) \delta \rho(w)$ with respect to the Boltzmann weight $\propto e^{-\mathcal{S}}$ is given by
\begin{equation}
\langle \delta \rho(z) \delta \rho(w) \rangle = G(z-w),
\end{equation}
where the Green's function satisfies
\begin{equation}
\int_{-a}^{a} dw \, K(x-w)G(w-z) = \delta(x-z).
\end{equation}
We refer to Refs. \cite{beenakker1993universality,ChalkerMacedo} for a careful justification of analogous arguments for the DMPK equation. To proceed further, we note that the bulk short-time behaviour described above is consistent with the linear approximation
\begin{equation}
\label{eq:linapprox}
K(z-w) \approx -|z-w|
\end{equation}
to the kernel $K$, for which the Green's function
\begin{equation}
\label{eq:greenapprox}
G(z-w) = -\frac{1}{2}\delta''(z-w)
\end{equation}
defines an inverse in the bulk region. We note that Eq. \eqref{eq:greenapprox} implies the results of Ref. \cite{flack2023exact} on the variance of linear statistics for the one-dimensional Coulomb gas, which is a nontrivial test of its validity. Let us define the mean moments $n^{(\alpha)} = \int_{-a}^{a} dz \, \bar{\rho}(z) \, e^{2\alpha z}$ and their fluctuations $\delta n^{(\alpha)} = \int_{-a}^{a} dz \, \delta \rho(z) \, e^{2\alpha z}$.  Combining the short-time approximations Eq. \eqref{eq:uniformansatz} and Eq. \eqref{eq:greenapprox} and neglecting boundary effects yields the estimates
\begin{equation}
 n^{(\alpha)}(t) \approx \frac{1}{\alpha \Gamma t}\exp{\left(\frac{\alpha N \Gamma t}{2}\right)}, \quad 
\langle \delta n^{(\alpha)}(t) \delta n^{(\beta)}(t) \rangle \approx -\frac{(\alpha-\beta)^2}{4(\alpha+\beta)}\exp{\left(\frac{(\alpha+\beta)N\Gamma t}{2}\right)}
\end{equation}
for $N \Gamma t \gg 1$. From these expressions, we deduce that the leading contribution to the covariance of R{\'e}nyi entropies grows quadratically in time
\begin{equation}
\label{eq:RenyiFluct}
\langle \delta S^{(\alpha)}(t) \delta S^{(\beta)}(t)\rangle \approx \frac{\alpha \beta (\Gamma t)^2}{4(\alpha-1)(\beta-1)} \left( \frac{(\beta-1)^2}{\beta+1} + \frac{(\alpha-1)^2}{\alpha+1}-\frac{(\alpha-\beta)^2}{\alpha+\beta}\right)
\end{equation}
for $N \Gamma t \gg 1$. In particular, since $(\Gamma t)^2 \ll |\log{\Gamma t}|$ for $\Gamma t \ll 1$, this guarantees that fluctuations in the R{\'e}nyi entropies will be small compared to the mean R{\'e}nyi entropies $\bar{S}^{(\alpha)}$ until the end of the short-time regime $\Gamma t \approx 1$, at which the $\bar{S}^{(\alpha)}$ are order one by Eq. \eqref{eq:weakshortrenyi} and thus have magnitude comparable to their fluctuations Eq. \eqref{eq:RenyiFluct}.

\subsection{Long times: log-normality, Lyapunov exponents and level repulsion}
\label{sec:longweak}
Anticipating that the $z_n$ are well-separated for $s \gg 1$, we define the asymptotic drift velocities 
\begin{equation}
\label{eq:driftvel}
c_n = \lim_{z_{j}-z_k \to \infty, \, j < k} D_n(\vec{z}) = 2(N+1-2n).
\end{equation}
Noting that $\sum_{n=1}^N c_n^2 = 4N(N^2-1)/3$ and $\sum_{j<k} (z_j-z_k) = \frac{1}{2}\sum_{n=1}^N c_n z_n$, we can rewrite Eq. \eqref{eq:exactpdf} as
\begin{equation}
\label{eq:lognpluscorr}
P(\vec{z},s) =  \frac{\prod_{j<k} (z_j - z_k)}{(4s)^{N(N-1)/2}\prod_{n=1}^{N-1}n!} \prod_{j<k} \left(1-e^{-2(z_j-z_k)}\right) \prod_{n=1}^N \frac{1}{\sqrt{4\pi s}}e^{-\frac{(z_n-c_ns)^2}{4s}}.
\end{equation}
Thus as $s \to \infty$, 
\begin{equation}
P(\vec{z},s) = P_{\mathrm{l.t.}}(\vec{z},s) + \mathcal{O}(s^{-1/2}),
\end{equation}
where the dominant contribution at asymptotically long times is given by
\begin{equation}
\label{eq:lognormalapprox}
P_{\mathrm{l.t.}}(\vec{z},s) =  \frac{\prod_{j<k} (c_j - c_k)s}{(4s)^{N(N-1)/2}\prod_{n=1}^{N-1}n!} \prod_{n=1}^N \frac{1}{\sqrt{4\pi s}}e^{-\frac{(z_n-c_ns)^2}{4s}} = \prod_{n=1}^N \frac{1}{\sqrt{4\pi s}}e^{-\frac{(z_n-c_ns)^2}{4s}},
\end{equation}
since the Vandermonde determinant
\begin{equation}
\prod_{j<k} (c_j - c_k)s = \prod_{j<k} 4(k-j)s = (4s)^{N(N-1)/2} \prod_{n=1}^{N-1}n!.
\end{equation}
We deduce that long-time asymptotic behaviour of $P(\vec{z},s)$ is log-normal.

We note that the drift velocities Eq. \eqref{eq:driftvel} correspond to Lyapunov exponents of $e^{z_n}$ as a function of the rescaled time $s$~\cite{Ipsen_2016}. Combining Eqs. \eqref{eq:lognormalapprox}, \eqref{eq:zinsigma} and \eqref{eq:defofs} then implies that $
\sigma_n(t) \sim e^{\lambda_n t}$ as $t \to \infty$, where the Lyapunov exponents for the singular values as a function of time are given by
\begin{equation}
\label{eq:ExactLyapWeak}
\lambda_n = -pL \log 2 + \frac{\Gamma}{8}(N+2-4n), \quad n =1,2,\ldots,N.
\end{equation}
Thus the Lyapunov spectrum for weak measurements is linear in $n$ for all $n$ (in contrast to the Lyapunov spectrum for projective measurements, Eq. \eqref{eq:ExactLyapProj}, which is only linear for $n \ll M$). In particular, we can read off the purification time defined by Eq. \eqref{eq:purificationtime}, which is given by
\begin{equation}
\tau_{\mathrm{P}} = \Gamma^{-1}
\end{equation}
for this model as expected.

Despite accurately capturing the behaviour of widely separated singular values, the asymptotic expression Eq. \eqref{eq:lognormalapprox} leads to unphysical predictions for the joint distribution function of $z_1$ and $z_2$, predicting for example that the mean ratio of the two largest singular values $\nu(t)=e^{2(z_2-z_1)}$ does not decay in time. This is because Eq. \eqref{eq:lognormalapprox} omits the level repulsion implied by our ordering of the $z_n$. To correct this, we make a more accurate long-time approximation to the exact distribution function Eq. \eqref{eq:exactpdf} that is again expected to hold at asymptotically long times, whereby subleading singular values $z_n$ with $n \geq 3$ are again treated as lognormal but the leading two singular values $z_1$ and $z_2$ are treated exactly. Integrating over $z_n$ for $n \geq 3$ then yields the ``improved'' joint distribution function
\begin{equation}
\tilde{P}_{\mathrm{l.t.}}(z_1,z_2,s) = \frac{(z_1-z_2)}{4s}\left(1-e^{-2(z_1-z_2)}\right) \frac{1}{4\pi s}e^{-\frac{(z_1-c_1s)^2+(z_2-c_2s)^2}{4s}}
\end{equation}
for the two leading singular values. Let us now confirm that this yields physically reasonable predictions for $\nu(t)$. 

It will be helpful to introduce centre-of-mass coordinates $\bar{z} = \frac{z_1+z_2}{2}$ and $\zeta = z_1 - z_2$, in which the improved joint distribution function
\begin{equation}
\tilde{P}_{\mathrm{l.t.}}(z_1,z_2,s) = A(\bar{z})B(\zeta)
\end{equation}
is separable, with
\begin{equation}
A(\bar{z}) = \frac{1}{\sqrt{2\pi s}} e^{-\frac{1}{2s}(\bar{z}-(2N-4)s)^2}   
\end{equation}
and
\begin{equation}
\label{eq:improvedmeasure}
B(\zeta) = \frac{1}{\sqrt{8\pi s}} \frac{\zeta}{4s} \left(e^{-\frac{1}{8s}(\zeta-4s)^2} - e^{-\frac{1}{8s}(\zeta+4s)^2} \right).
\end{equation}
The latter manifestly exhibits level repulsion as $\zeta \to 0^+$. With respect to the improved long-time measure Eq. \eqref{eq:improvedmeasure}, typical values of $\nu(t)$ behave as
\begin{equation}
\label{eq:lognu}
\mathbb{E}[\log{\nu}(t)] \sim -\Gamma t, \quad \Gamma t \gg 1,
\end{equation}
implying that
\begin{equation}
\label{eq:weaklongrenyi}
\mathbb{E}[\log{S_{\mathbf{m}}^{(\alpha)}(t)}] \sim \begin{cases}
-\Gamma t, & \alpha \geq 1, \\
-\alpha \Gamma t, & 0<\alpha < 1,
\end{cases}
\end{equation}
which recovers Eqs. \eqref{eq:lognu} and \eqref{eq:weaklongrenyi} at times $t \gg \Gamma^{-1}$, while the mean value of $\nu(t)$ also decays exponentially,
\begin{equation}
\mathbb{E}[\nu(t)] \sim \frac{16}{9} \frac{1}{\sqrt{\pi(\Gamma t)^3}} e^{-\Gamma t/4}, \quad \Gamma t \gg 1,
\end{equation}
(albeit at a slightly different rate from the typical value), implying for example that the mean $\alpha>1$ R{\'e}nyi entropies  
\begin{equation}
\mathbb{E}[S_{\mathbf{m}}^{(\alpha)}(t)] \sim \frac{\alpha}{\alpha-1} \frac{16}{9} \frac{1}{\sqrt{\pi(\Gamma t)^3}} e^{-\Gamma t/4}, \quad \alpha > 1,
\end{equation}
decay exponentially for $t \gg \Gamma^{-1}$.

\section{Conclusion}
\label{sec:concl}

We have presented various solvable models of monitored quantum circuits (``monitored Haar-random quantum dots'') that realize the entangling phase of monitored quantum dynamics. While these models have antecedents in the literature~\cite{nahum2021measurement,fidkowski2021dynamical,schomerus2022noisy}, the full extent of their analytical tractability does not appear to have been exploited until now. This analytical tractability has allowed us to derive the first exact expressions that we are aware of for quantities such as the purification time, the Lyapunov spectrum and the distribution of Born probabilities in the entangling phase of a monitored quantum system. By constructing explicit mappings from monitored Haar-random quantum dots to well-studied models in random matrix theory, such as products of truncated unitary matrices~\cite{zyczkowski2000truncations,Akemann_2014} and isotropic Brownian motion~\cite{le1985isotropic,Ipsen_2016}, we have further provided a template for realizing random-matrix universality in monitored quantum systems. 

Our proposed notion of universality for such systems is analogous to the use of random matrix theory (RMT) as a description of the spectra of closed quantum systems; in the same way that RMT captures spectral correlations of generic chaotic quantum systems, we conjecture that the models discussed in this paper capture certain features of the entangling phase of generic monitored quantum systems, including the spatially local quantum circuits in which the entangling phase was first identified~\cite{LiChenFisher,SkinnerRuhmanNahum}.

Before describing our proposal in detail, let us explain how universality for monitored quantum systems must differ from random-matrix universality and the related eigenstate thermalization hypothesis (ETH)~\cite{ETHrev} in closed quantum systems. Most obviously, the qualitative behaviour of the unstructured models considered in this paper depends on three parameters: the Hilbert space dimension $N$, the purification time $\tau_{\mathrm{P}}$ and the time $t$ under consideration. The latter two parameters have no natural counterpart in closed quantum systems, whose dynamics is Hamiltonian and therefore time-translation invariant. In particular, we expect that monitored quantum systems will exhibit different facets of universal behaviour depending on the dimensionless parameter $t/\tau_{\mathrm{P}}$, with qualitative differences between the short-time ($t / \tau_{\mathrm{P}} \ll 1$) and long-time ($t / \tau_{\mathrm{P}} \gg 1$) regimes. Second, instead of the matrix elements of local observables and energy-level-spacing statistics that are usually of primary interest in formulating and testing ETH~\cite{ETHrev}, the monitored setting suggests new arenas for universality, such as the statistics of single-trajectory Born probabilities and R{\'e}nyi entropies discussed in this paper, both of which reflect the singular-value statistics of the underlying Kraus operators.

First consider the short-time regime, $t/\tau_{\mathrm{P}} \ll 1$. Our results on the distribution of Born probabilities in Section \ref{sec:born} and on the sample-to-sample fluctuations of R{\'e}nyi entropies in Section \ref{sec:shortweak} suggest that for the models studied in this paper, both the Born probabilities and the R{\'e}nyi entropies will be narrowly distributed about their means (possibly on a log scale) in the short-time regime. We expect this conclusion to hold more generally for the entangling phase in arbitrary spatially local models. This means that on timescales that are short compared to the purification time (excluding initial transients as in Section \ref{sec:veryshort}), the behaviour of the system is to a first approximation decoupled from its measurement history, and to this extent, all quantum trajectories are statistically alike. 

By contrast, our results on both Born probabilities and R{\'e}nyi entropies at long times (Sections \ref{sec:born}, \ref{sec:longproj} and \ref{sec:longweak}) indicate that these quantities exhibit a broad and normal (or normal-derivative-like) distribution on a log scale at long times $t/\tau_{\mathrm{P}} \gg 1$. Thus the observed behaviour in the long-time regime will depend strongly on the measurement history, and distinct quantum trajectories will exhibit very different properties. We expect similar behaviour for the entangling phase of spatially local systems at long times, even at the level of the shapes of these probability distributions, which should reflect universal properties of large products of identically distributed random matrices~\cite{Akemann_2019}.

Finally, we note that the state-vector of the system at long times should exhibit some degree of universality by Eq. \eqref{eq:purifiedstate}, which predicts that the long-time density matrix along any given trajectory is a random pure state. This is reminiscent of Berry's conjecture~\cite{berry1977regular}, which posits that the eigenvectors of a chaotic Hamiltonian are distributed as Gaussian random vectors.
For the projectively-measured Haar-random quantum dot, the long-time state-vector $\mathbf{v}_1$ in Eq. \eqref{eq:purificationtime} is distributed as a Gaussian random vector in the $M$-dimensional image of the most recent measurement layer. More generally, for spatially local systems the long-time state-vector $\mathbf{v}_1$ is no longer expected to be perfectly Gaussian for entangling phases, as can be seen from the presence of logarithmic in $L$ corrections to the bipartite entanglement entropy of the long-time state in spatially local systems~\cite{LiEnt,FanEnt}, which would otherwise exhibit purely volume-law dependence on $L$ as predicted by the Page curve~\cite{Page}.

Note that in this paper, we do not consider the Born-rule weighted averages of quantities such as R{\'e}nyi entropies. 
We emphasise that the possibility of Born-rule weighting does not arise for the Lyapunov exponents, which by definition characterize a product of independently and identically distributed random matrices. On the other hand, for R{\'e}nyi entropies we expect that introducing Born-rule averaging will leave our results essentially unchanged at times $t \ll \tau_{\mathrm{P}}$ but will have an effect at times $t \gg \tau_{\mathrm{P}}$. Accounting for Born-rule averaging within the approach in Section \ref{sec:secweak} would lead to a distinct Fokker-Planck equation from the one studied above, as discussed in Ref. \cite{fidkowski2021dynamical}.

Important goals for future work include directly testing the above predictions of universality in spatially local realizations of the entangling dynamical phase (for which there is currently a dearth of analytical understanding) and developing a similarly universal characterization of disentangling phases. Another interesting question is whether the kinds of quantities that we have computed analytically in this paper, such as Lyapunov spectra and distributions of Born probabilities, can shed further light on monitored phases of ``non-interacting" or Gaussian quantum circuits~\cite{nahum2021measurement,fidkowski2021dynamical,Cao_2019,Nahum_2020,chen2020emergent,Alberton_2021}, for which it is known that the purification time scales quadratically with the system size~\cite{fidkowski2021dynamical,nahum2021measurement}. In the Haar-random setting, it seems worth understanding how far the generalizations of the Porter-Thomas distribution that we identify in Section \ref{sec:born} imply hardness-of-sampling results analogous to what is known for random unitary circuits~\cite{Bouland_2018}. Such results would both complement the prediction of a computational-complexity transition in monitored random circuits~\cite{suzuki2023quantum} and
lend theoretical support to proposals~\cite{LiCrossEnt,garratt2023probing} for diagnosing monitored dynamical phases that avoid post-selection by computing cross-entropies instead.

Shortly after this work was completed, related results appeared in two papers \cite{deluca2023universality,gerbino2024dyson}.

\section*{Acknowledgments}
We thank B. Fefferman, S. Gopalakrishnan, D.A. Huse, A. Nahum and A. Vishwanath for helpful discussions and V. Khemani especially for an early conversation setting up some of the questions addressed in this paper.  This research was supported in part by the National Science Foundation under Grant No. NSF PHY-1748958 at KITP. V.B.B. was supported by a fellowship at the Princeton Center for Theoretical Science during part of the completion of this work. J.T.C. was supported in part by EPSRC Grant EP/S020527/1. S.L.S. was supported by a Leverhulme Trust International Professorship, Grant Number LIP-202-014. For the purpose of Open Access, the authors have applied a CC BY public copyright license to any Author Accepted Manuscript version arising from this submission. 


\bibliography{biblio.bib}
\end{document}